\documentclass[
  twoside,
  10pt
]{amsart}
\RequirePackage{amsaddr}

\newcommand*{\dfFix}{final} 
\newcommand*{\dfKey}{final} 
\RequirePackage{datetime}
\settimeformat{xxivtime}

\newdateformat{dmydate}{\THEDAY~\monthname[\THEMONTH]\ \THEYEAR}
\newdateformat{ndmydate}{\twodigit{\THEDAY}.\twodigit{\THEMONTH}.\THEYEAR}
\dmydate

\RequirePackage[T1]{fontenc}
\RequirePackage{amsmath}
\RequirePackage{amssymb}
\RequirePackage{microtype}

\RequirePackage{booktabs}
\RequirePackage{graphicx}
\graphicspath{{./figures/}{../figures/}}
\RequirePackage{hyperref}

\RequirePackage{siunitx}
\newcommand*{\murad}{\micro\radian}
\newcommand*{\mrad}{\milli\radian}

\RequirePackage{xcolor}
\definecolor{hrBlue}{cmyk}{1.00, 0.25, 0.25, 0.00}
\definecolor{hrRed}{cmyk}{0.25, 1.00, 1.00, 0.00}
\definecolor{hrGreen}{cmyk}{1.00, 0.25, 1.00, 0.00}
\definecolor{Grey}{gray}{0.60}

\hypersetup{%
  colorlinks,
  pdfborder = {0 0 0},
  bookmarksdepth = section,
  hyperfootnotes = false,
  citecolor = hrGreen,
  linkcolor = hrRed,
  urlcolor = hrBlue 
}

\RequirePackage[numbers]{natbib}
\DeclareUrlCommand\Doi{\urlstyle{rm}}
\providecommand{\doi}[1]{\href{http://dx.doi.org/#1}{%
                         \textsc{doi:}\,\Doi{#1}}}

\RequirePackage{paralist}

\RequirePackage[\dfFix,nomargin,inline,index]{fixme}

\RequirePackage[color,\dfKey]{showkeys}
\definecolor{refkey}{rgb}{.8,.8,1}
\definecolor{labelkey}{rgb}{.5,.5,1}

\newcommand{\hairsp}{\hspace{0.3pt}}
\newcommand{\ie}{\emph{\mbox{i.\hairsp{}e.}}}
\newcommand{\eg}{\emph{\mbox{e.\hairsp{}g.}}}
\newcommand{\cf}{\emph{cf.}}
\newcommand{\etc}{\emph{etc.}}
\newcommand{\teapot}{\textsc{TEAPOT}}

\newcommand{\gpuST}{\textsc{gpuSpinTrack}}

\newcommand{\ual}{\textsc{UAL}}
\newcommand{\cpu}{\textsc{CPU}}
\newcommand{\gpu}{\textsc{GPU}}
\newcommand{\dkd}{\textsc{DKD}}
\newcommand{\dks}{\textsc{DKSKD}}
\newcommand{\dkds}{\textsc{DKDSDKD}}
\newcommand{\bkmk}{\textsc{BK/MK}}
\newcommand{\pwc}{\textsc{PWC}}
\newcommand{\ir}{\textsc{IR}}
\newcommand{\rhic}{\textsc{RHIC}}
\newcommand{\tbmt}{Thomas-\textsc{BMT}}
\newcommand{\isf}{\textsc{ISF}}
\newcommand{\edm}{\textsc{EDM}}
\newcommand{\nersc}{\textsc{NERSC}}
\newcommand\fake[1]{}

\newcommand{\mhsp}{\mspace{2mu}}
\newcommand{\ud}{\mathop{}\!\mathrm{d}}
\newcommand{\ui}{\mathrm{i}}
\newcommand{\ue}{\mathrm{e}}
\newcommand{\dd}[3][]{\ud^{#1}{#2}/\!\ud{#3}^{#1}}
\newcommand{\Dd}[3][]{\frac{\ud^{#1}{#2}}{\ud{#3}^{#1}}}

\newcommand{\Dt}[2][]{\frac{\ud^{#1}{#2}}{\ud{t}^{#1}}}

\newcommand{\ptDd}[3][]{\frac{\partial^{#1}{#2}}{\partial{#3}^{#1}}}

\newcommand{\uV}[1]{\hat{#1}}
\newcommand{\V}[1]{\vec{#1}}
\newcommand{\spin}{\V{\mathcal{S}}}
\newcommand{\bigOmega}{\V{\Omega}}
\newcommand{\uhat}{\uV{u}}
\newcommand{\xhat}{\uV{x}}
\newcommand{\yhat}{\uV{y}}
\newcommand{\zhat}{\uV{z}}
\newcommand{\sL}{^\text{(L)}}
\newcommand{\sNL}{^\text{(NL)}}
\DeclareMathOperator{\re}{Re}

\DeclareMathOperator{\liecolon}{:}
\newcommand{\lieop}[1]{\liecolon\!{#1}\!\liecolon}

\newcommand{\pfrac}[2]{\genfrac{(}{)}{}{}{#1}{#2}}
\newcommand*{\brho}{(B\rho)_\circ}
\newcommand*{\frakb}{\mathfrak{b}}
\newcommand*{\po}{p_\circ}
\newcommand*{\bo}{\beta_\circ}
\newcommand*{\go}{\gamma_\circ}

\newcommand*{\px}{P_x}
\newcommand*{\py}{P_y}
\newcommand*{\pt}{P_t}
\newcommand*{\ps}{P_s}

\newcommand*{\refS}[1]{section~\ref{sec:#1}}
\newcommand*{\refSs}[1]{sections~\ref{sec:#1}}
\newcommand*{\refE}[1]{\eqref{eq:#1}}

\newcommand*{\refF}[1]{figure~\ref{fig:#1}}
\newcommand*{\refFc}[1]{Figure~\ref{fig:#1}}
\newcommand*{\refFs}[1]{figures~\ref{fig:#1}}
\newcommand*{\refFsc}[1]{Figures~\ref{fig:#1}}

\begin{document}

\title{Accurate and efficient spin integration
       for particle accelerators}
\author[D.\hairsp{}T. Abell, D. Meiser]{Dan T. Abell and Dominic Meiser}
\address{\vspace{-1.0ex}
  Tech-X Corporation,
  5621 Arapahoe Avenue,
  Boulder, Colorado 80303, USA}
\author[V.\hairsp{}H. Ranjbar]{Vahid H. Ranjbar}
\address{\vspace{-1.0ex}
  Brookhaven National Laboratory,
  Upton, New York 11973 USA}
\author[D.\hairsp{}P. Barber]{Desmond P. Barber$^\star$}
\address{\vspace{-1.0ex}
  Deutsches Elektronen-Synchroton (DESY),
  22603 Hamburg, Germany}
\thanks{$^\star$Also Visiting Staff Member at the Cockcroft Institute,
Sci-Tech Daresbury, UK, and at the University of Liverpool, UK.}

\date{13 January 2015}

\begin{abstract}
Accurate spin tracking is a valuable tool for understanding
spin dynamics in particle accelerators and can help improve the
performance of an accelerator. In this paper, we present a detailed
discussion of the integrators in the spin tracking code \gpuST.
We have implemented orbital integrators based on drift-kick,
bend-kick, and matrix-kick splits. On top of the
orbital integrators, we have implemented various integrators for
the spin motion. These integrators use quaternions and Romberg
quadratures to accelerate both the computation and the convergence
of spin rotations. We evaluate their performance and accuracy
in quantitative detail for individual elements as well as for
the entire \rhic\ lattice.
We exploit the inherently data-parallel nature of spin tracking
to accelerate our algorithms on graphics processing units.
\end{abstract}

\maketitle

\section{Introduction}
\label{sec:intro}

The origin of nucleon spin remains an enduring puzzle in nuclear
physics%
  ~\cite{Aidala:2013:SpinStructNucleon},
and elucidating this puzzle is the principal focus of polarized
beam experiments at the Relativistic Heavy Ion Collider (\rhic)
at Brookhaven National Laboratory%
  ~\cite{Harrison:2002:RHICaccelerator,%
         Bai:2008:RHICStatusPP,%
         Huang:2011:RHICStatus}.
Because statistical uncertainties scale inversely with the square
of the polarization%
  ~\cite{Fischer:2012:Impact3DPol},
optimizing the polarization is essential for the efficient use
of experimental resources.

Computer simulations serve an important r\^ole in understanding
and improving beam polarization. For example, the invariant spin
field (\isf) places an important upper bound on the maximum
attainable polarization of a stored beam%
  ~\cite{Heinemann:1996:StrobAvg,%
         Hoffstaetter:2006:HEPPbeams},
and hence a knowledge of the \isf\ and how it varies with the
machine optics is essential to optimizing the beam polarization.

Another motivation for spin-orbit tracking sim\-u\-la\-tions%
---especially of high ac\-cu\-ra\-cy---derives from proposals to use
storage rings in searches for a permanent electric dipole moment
(\edm) in protons and deuterons%
  ~\cite{Onderwater:2011:SearchEDM}.
Assessing the sensitivity of such experiments will require
long-term spin-orbit simulations of unprecedented accuracy%
  ~\cite{Lehrach:2012:CompNeedsEDM}.
Other projects that can benefit from similar studies include
MEIC~\cite{Accardi:2011:NucPhysMEIC,%
           JLab:2012:SciReqEIC},
the LHeC~\cite{LHeC:2012:PhysicsAndDesign,%
               Bruning:2013:LHeC},
and the muon $g-2$ experiment%
  ~\cite{Miller:2007:MuonGMinus2,%
         Roberts:2010:StatusFermilabGMinus2}.
In addition, accurate spin-orbit tracking enables one to perform
careful tests of mathematical concepts related to spin dynamics%
  ~\cite{Barber:2004:Qperiodic}.

In spite of the manifest need for accurate simulations of
spin dynamics, relatively few efforts have been made to develop
codes that include spin tracking. In the context of hadron machines,
which is the context of this paper, spin-orbit tracking codes for
large storage rings and accelerators first emerged in the 1990s
in response to the needs of specific projects.
M\'eot added spin tracking to \textsc{Zgoubi} in 1992%
  ~\cite{Meot:1992:SpinZgoubi, Meot:Zgoubi}.
Hoffst\"atter and Berz did the same for \textsc{COSY-Infinity} in 1993%
  ~\cite{Berz:COSYInfinity}.
Luccio developed \textsc{SPINK} in the mid-1990s to support the goal of
accelerating polarized protons in \rhic%
  ~\cite{Luccio:1995:SimCodeSPINK, Luccio:1999:Spink}.
Hoffst\"atter and Vogt created \textsc{SPRINT} to study the feasibility
of attaining proton polarization at very high energy in HERA%
  ~\cite{Vogt:2000:PhDthesis, Hoffstaetter:2006:HEPPbeams}.
The code \textsc{FORGET-ME-NOT}, by Golubeva and Balandin, was developed
for this purpose also%
  ~\cite{Balandin:1997:HamMethSpin}.
In the mid-2000s spin motion was added to \textsc{PTC} by Forest%
  ~\cite{Forest:2002:PTC, Forest:2006:FPP.PTC,
         Forest:2010:OrbitalSpinFPP}
and to \textsc{Bmad} by Sagan%
  ~\cite{Sagan:BMAD}.
Mane has recently added the code \textsc{ELIMS} to this list%
  ~\cite{Mane:2012:CalcHOSpinRes}.


Some of these codes also include effects particular to electrons%
---especially synchrotron radiation. In addition, there are other
more specialized codes that handle electrons: Of signal importance
is the development of \textsc{SLIM} by Chao%
  ~\cite{Chao:1981:EvalRadSpinPol}
in the early 1980s.


The original version of \textsc{SPINK} used orbit transport matrices
generated by \textsc{MAD8}---later by \textsc{MAD-X}---to compute the orbital
motion. That version was used for extensive studies of the
beam polarization in \rhic%
  ~\cite{Luccio:1995:SimCodeSPINK}.
\textsc{SPINK} was later incorporated into the UAL framework and modified
to use \textsc{TEAPOT}'s orbital integrators. The current work arose out
of an effort to port \textsc{UAL-SPINK} to \gpu\ platforms. In the course
of that work, however, it was discovered that even when using
\textsc{TEAPOT}'s orbital integrators, the code had difficulties with
spin convergence%
  ~\cite{Ranjbar:2011:SpinTrackGPUs,Abell:2013:GPUAccelSpin},
especially in the neighborhood of a strong spin resonance.
Addressing this issue meant slowing down what were already
numerically-demanding spin tracking simulations.

Here we present a very accurate and efficient spin-orbit tracking
code, \gpuST, that we have incorporated into the \ual\ framework%
  ~\cite{Malitsky:TowardsOpenSourceUAL}.
Because we have found that the accuracy of the orbital data
has a significant impact on the accuracy of the spin tracking, our
code relies on first performing very accurate symplectic integration
for the orbital motion. With orbital data in hand, we then use the
Thomas-Bargman-Michel-Telegdi (\tbmt, or \textsc{T-BMT}) equation%
  ~\cite{Thomas:1927:KinElectronAxis,%
         Bargmann:1959:PrecessionPP}
to integrate the spin motion.
Our integration methods include the effects of acceleration on
both orbit and spin degrees of freedom. We note that because
\gpuST\ tracks the full nonlinear orbital motion and the full
3D spin motion, it is sensitive to the full range of spin-orbit
resonances. This is particularly important in the context of
acceleration across spin-orbit resonances. The most significant
aspect of this work is that we have found a means of accelerating
the convergence of the spin integration.

To obtain reliable results from numerical studies, one must have
a quantitative understanding of the errors inherent in a given
simulation.  With a quantitative model for the errors, one can
perform simulations of the desired accuracy without wasting
computational effort motivated by overly pessimistic error estimates.
Another significant contribution of this paper is a detailed
analysis of the accuracy of our integrators. We study how
the orbital and spin motion converge after one turn, and
also the long-time stability of our algorithms. For the spin
dynamics, we also examine the errors across individual elements.

Numerical efficiency is another important consideration for spin-orbit
tracking codes. To increase the speed of our simulations, we have
implemented all our integrators on a graphics processing unit (\gpu).
The embarrassingly parallel nature of spin-orbit tracking (in the
absence of space-charge) makes this type of computation an ideal fit
to the highly parallel architecture of \gpu{}s.

In the next three sections, we describe the dynamical model used
for our simulations (\refS{dynModel}), the orbital integrators
we implement (\refS{orbitInteg}), and our approach to spin
integration (\refS{spinInteg}). The latter section describes
what we refer to as \emph{Romberg quadratures for spin}, our
method for accelerating the convergence of spin integration.
In \refS{performance}, we describe the performance of both
orbit and spin integrators. We give a brief description of the
\gpu\ implementation in \refS{GPUimplementation}. Finally, in
\refS{conclusion}, we conclude by discussing our results.

\section{The Dynamical Model}
\label{sec:dynModel}

In this study, we ignore the effects of synchrotron radiation and
space-charge forces. We therefore model the orbital dynamics in an
accelerator using a single-particle Hamiltonian appropriate to the
externally applied magnetic and electric fields of a particle accelerator.
For the spin dynamics, we treat a particle's spin expectation value
as a unit three vector%
  \footnote{Note that for particles with integer spin, there can
            be a state with vanishing spin expectation value.
            But this state is usually of little interest for
            experiments with polarized beams.}
that obeys the \tbmt\ equation%
  ~\cite{Thomas:1927:KinElectronAxis,%
         Bargmann:1959:PrecessionPP}.

\subsection{Model for orbital dynamics}
\label{sec:orbitModel}

To describe the orbital motion of particles in a beamline element,
we use longitudinal distance $s$ along the element as our
independent variable, and we use the same canonical phase-space
co\"ordinates as MAD~\cite{CERN:2002:MadXmanual}:
\begin{equation}\label{eq:psvars}
  \V z = (X, \px, Y, \py, T, \pt).
\end{equation}
Here $X$ and $Y$ denote the horizontal and vertical co\"ordinates
in the local reference frame of our beamline element; $\px$ and
$\py$ denote the corresponding conjugate momenta, divided by a fixed
\emph{scale momentum} $\po = m\go\bo c$; $T = -c\Delta t$ measures
the flight time (times the speed of light, $c$) relative to a reference
particle with momentum $\po$; and $\pt$ denotes the energy deviation
scaled by $\po c$, so that
\begin{equation}\label{eq:pt}
  \pt = \frac{m\gamma c^2 - m\go c^2}{\po c}
      = \frac{m\gamma c^2}{\po c} - \frac{1}{\bo}.
\end{equation}
Note that the minus sign in the definition of $T$ means that a
\emph{positive} value for $T$ implies that our particle arrives
\emph{earlier} than the reference particle.

For magneto-static elements with strictly transverse fields that
do not vary along the length, our Hamiltonian for a particle of
charge $q$ has the general form
\begin{equation}\label{eq:generalH}
  H = -(1 + hX)
       \sqrt{1 + \frac{2}{\smash{\bo}}\pt + \pt^2 - \px^2 - \py^2}
      - \frac{q}{\po}(1 + hX) A_s(X,Y) + \frac{1}{\bo}\pt,
\end{equation}
where $h = 1/\rho_c$ denotes the curvature of the local co\"ordinate
frame---$h$ will vanish except in sector bends---and $A_s(X,Y)$
denotes the longitudinal component of the element's vector potential.
(We choose a gauge in which the transverse components of the vector
potential vanish.) Note that this Hamiltonian assumes any bending
occurs only in the horizontal plane; for bends in any other plane,
we simply apply rotations before and after a horizontal bend. The
last term, $\pt/\bo$, accounts for the flight time of a particle
with the reference momentum traversing the orbit $X=Y=0$.

When integrating through a dipole, one must exercise some care
concerning the relation between the fixed geometry, defined by
a magnet's curvature and placement, and the variable physics,
defined by a magnet's field strength. For sector bends, the
fixed geometry means the curvature $h$ of the local frame is
fixed. In the usual practice, one then sets the scale momentum
$\po$ according to this curvature, so that $\po=qB_1\rho_c$,
where $B_1$ denotes the magnet's design field strength. Then
$qB_1/\po=h$, and the vector potential term in \refE{generalH}
simplifies to $\frac{1}{2}(1+hX)^2$, but \emph{only} if the
magnet is correctly powered. To include the possibility of
mispowering errors, one must not thus confuse the geometry
with the physics; one must instead retain the dependence of the
Hamiltonian on the actual magnetic field strength.

For rectangular bends, which have curvature $h=0$, a more
fundamental difficulty arises from the fact that the orbital
motion is most simply integrated using Cartesian co\"ordinates,
whereas the curved design orbit does not follow the straight
magnet axis. This means the horizontal phase-space variables
$(X,\px)$, which describe deviations from the magnet
axis, cannot describe deviations from any choice of
reference orbit. Moreover, for a given fixed bending angle,
the path length---hence also the flight time---will depend on
the entrance angle. As a consequence, the flight time along
even a reference orbit will depend on how the rectangular bend
is oriented relative to adjacent beamline elements. If we wish
the phase-space variables of \refE{psvars} at both entrance
and exit of this magnet to represent deviations, then we must
view the Hamiltonian \refE{generalH} as written in a set of
variables \emph{internal} to this magnet (and we must drop the
term $\pt/\bo$). We then perform the appropriate conversions
at entrance and exit.

Of course real magnets, because they obey the dictates of Maxwell,
do \emph{not} have ``strictly transverse fields that do not
vary along the length''. That fiction---to which no sensible
spectroscopist subscribes---proves useful in much of accelerator
physics because (a) long magnets have fields for which that fiction
is close to the truth, (b) short magnets yield orbital motion
that is dominated by the integrated strength, and (c) fringe-field
effects in multipole magnets, which arise primarily from the
longitudinal field component, have contributions from both entrance
and exit that tend to cancel, particularly in short magnets%
  ~\cite{Forest:2006:GeomIntegPA}.
Still the field variation across a magnet fringe \emph{can} have
a significant impact on the beam dynamics. Cases where this is
especially true include solenoids, whose longitudinal fields
naturally have extended fringe regions; and rectangular bends,
for which non-normal entrance and exit angles lead to vertical
focusing%
  ~\cite{Brown:1982:SLAC75}.
In addition, nonlinear contributions in, for example, quadrupole
fringe fields can cause large-emittance beams to occupy a large
footprint in tune space%
  ~\cite{Papaphilipou:2000:MagCorrSNS}.

In this paper we shall, for the most part, avoid detailed
discussions of fringe fields. We make this choice for a couple
of reasons.
(i)~The simulations shown in this paper were all done in the
context of the Relativistic Heavy Ion Collider (\rhic) at
Brookhaven National Lab, for which fringe field effects are
relatively small.
(ii)~The subject of transfer maps for magnetic fringe fields is
complicated, and very good discussions are available elsewhere.
In addition to the work cited above, the reader may consult Forest
and colleagues%
  ~\cite{Forest:1988:HardEdgeFringe,Forest:1998:BeamDyn,%
         Forest:2006:FringeEffects1,Forest:2006:FringeEffects2}
for work done mostly in the context of a hard-edge limit,
or Dragt and colleagues%
  ~\cite{Venturini:1999:AccurCompXfer,Venturini:1999:MagXferMap,
         Mitchell:2010:AccXferMaps,Dragt:2015:LieMethNLDyn}
for work on transfer maps across realistic magnets that include
fringe fields.

\subsection{Model for spin dynamics}
\label{sec:spinModel}

We describe a particle's spin expectation value by a unit three
vector $\spin$. The precession of this spin in a magnetic field
$\vec{B}$ is governed by the \tbmt\ equation%
  ~\cite{Thomas:1927:KinElectronAxis,%
         Bargmann:1959:PrecessionPP},
which says that in the rest frame of the magnet---our laboratory
frame---the spin precesses according to the rule
\begin{equation}\label{eq:t-bmt.t}
  \Dt{\spin} = \spin \times \frac{q}{m\gamma}
    \bigl[(1 + G\gamma)\V{B}
          - G(\gamma - 1)(\uhat\cdot\V{B})\uhat\bigr].
\end{equation}
An additional term, not shown here, must be included in the
presence of electric fields.
In this equation, $G$ denotes the gyro-magnetic anomaly,
$(g-2)/2$, with $g$ the particle's gyromagnetic ratio; and
$\uhat$ denotes the unit velocity vector, obtained by normalising
the momentum vector $(\px,\py,\ps)^T$. Here $\ps$ denotes the
longitudinal component of the scaled momentum, given by
\begin{equation}\label{eq:ps}
    \ps = \sqrt{1 + \frac{2}{\smash{\bo}}\pt + \pt^2 - \px^2 - \py^2}.
\end{equation}

As in the case of orbital motion, it proves convenient to transform
the equation of spin motion \refE{t-bmt.t} to one with $s$ as
the independent variable. To do this, we multiply both sides of
\refE{t-bmt.t} by $\dd{t}{s} = 1/\dot{s}$. It is important to note
that in a curved Frenet-Serret co\"ordinate system,
$\dot{s} \neq v_s$. Rather, simple geometry tells us that
\[
  v_s = (\rho_c + X)\dot\theta = (1 + h X)\rho_c\dot\theta
      = (1 + h X)\dot s.
\]
Using this result, we now obtain the modified \textsc{T-BMT} equation
in the form
\begin{subequations}\label{eq:t-bmt.s}
\begin{equation}\label{eq:t-bmt.sx}
  \Dd{\spin}{s} = \bigOmega \times \spin,
\end{equation}
where
\begin{equation}\label{eq:t-bmt.so}
  \bigOmega = -\frac{1+hX}{\brho\ps}
    \bigl[(1+G\gamma)\vec{B} - G(\gamma-1)(\uhat\cdot\vec{B})\uhat\bigr]
    +h\hat{y},
\end{equation}
\end{subequations}
and $\brho=\po/q$ denotes the (signed) reference value of the magnetic
rigidity. The last term, $h\yhat$, accounts for the local frame
rotation (assuming the rotation is about the axis $\hat{y}$).

Our model does not include spin-dependent contributions to the orbital
motion, so-called Stern-Gerlach forces, because these are completely
negligible for essentially all
accelerators~\cite{Heinemann:1996:OnSternGerlach}.

\section{Orbit Integration}
\label{sec:orbitInteg}

To integrate a particle's orbital motion through a given beamline
element, we use the standard technique of writing the relevant
Hamiltonian as a sum of one or several integrable pieces. One then
obtains an approximate solution for the total Hamiltonian by an
appropriate concatenation of exact solutions for the separate pieces.
An important advantage of this approach of splitting the Hamiltonian
is that it produces a symplectic integrator---\ie\ an integrator
that preserves the fundamental structure that underlies any Hamiltonian
system. A second advantage is that by composing the partial solutions
in a symmetric fashion, one can achieve second-order convergence to
the exact solution. Indeed, more sophisticated symmetric compositions
allow one to achieve even higher order convergence%
  ~\cite{Yoshida:1990:ConstrHiOrder,%
         Forest:2006:GeomIntegPA}.
For an extensive and valuable introduction to the literature on such
\emph{splitting methods}, see McLachlan and Quispel%
  ~\cite{McLachlan:2002:SplitMeth}.

There are several ways one might split the Hamiltonian~\refE{generalH}.
One of the oldest---still widely used---is the drift-kick split that is
the basis of \teapot%
  ~\cite{Schachinger:1985:TEAPOT,%
         Forest:2006:GeomIntegPA}.
One virtue of the drift-kick split is that it applies to a wide
variety of beamline elements. If, however, we focus on particular
elements, we can tailor the split accordingly. For ideal bending
magnets---sector bends and rectangular bends---one can derive
exact solutions, see sections~\ref{sec:sbend} and \ref{sec:rbend}.
For quadrupoles, one can split off the linear transverse motion,
leaving a much smaller nonlinear kick as a correction, see
\refS{quad}. For higher-order multipoles, we return to the
drift-kick split of \teapot.

\subsection{Sector bend}
\label{sec:sbend}

For a pure sector bend, one may write the vector potential term
in \refE{generalH} as
\begin{equation}\label{eq:sbendA}
  -\frac{q}{\po} A_s(X,Y) = \frac{b_1}{2h\brho}(1+hX)^2,
\end{equation}
where $b_1$ denotes the actual dipole field strength, and $h$ denotes
the (fixed) physical curvature $1/\rho_c$ of the magnet. One can solve
the corresponding equations of motion analytically%
  ~\cite{Forest:1998:BeamDyn}.
First define the scaled momentum in the horizontal plane,
\begin{equation}\label{eq:planarP}
  P_\alpha = \sqrt{1 + \frac{2}{\smash{\bo}}\pt + \pt^2 - \py^2}.
\end{equation}
This quantity is a constant of the motion. Also define the dimensionless
parameter
\begin{equation}\label{eq:eta}
  \eta = \frac{b_1}{h\brho} = \frac{b_1/h}{\po/q},
\end{equation}
which measures the magnetic field strength relative to its design
value. Then, setting $s=0$ at the magnet entrance, we obtain the
sector bend trajectory in the form
\begin{subequations}\label{eq:sbend}
\allowdisplaybreaks
\begin{align}
  X(s)   &= \frac{1}{h}\left[ \frac{1}{\eta}\Bigl(
              \sqrt{P_\alpha^2 - \px^2(s)}
              - \frac{1}{h} \px^\prime(s)
              \Bigr) - 1 \right],\\
  \px(s) &= \px^i\cos(hs)
           + \bigl[\sqrt{P_\alpha^2 - (\px^i)^2}
                   - \eta(1 + h X^i)\bigr]\sin(hs),\\
  Y(s)   &= Y^i + \frac{1}{\eta}\py s + \frac{1}{\eta h}\py
              \left[\sin^{-1}\!\pfrac{\px^i}{P_\alpha}
                    - \sin^{-1}\!\pfrac{\px(s)}{P_\alpha}\right],\\
  \py(s) &= \py^i,\\
  \begin{split}
  T(s)   &= T^i - \frac{1}{\eta}\pt s
             - \frac{1}{\eta h}
               \biggl(\frac{1}{\beta_0} + \pt\biggr)
                \left[\sin^{-1}\!\pfrac{\px^i}{P_\alpha}
                    - \sin^{-1}\!\pfrac{\px(s)}{P_\alpha}\right] \\
         &\hphantom{= T^i\ } + \biggl(1 - \frac{1}{\eta}\biggr)\frac{s}{\bo},
  \end{split}\\
  \pt(s) &= \pt^i,
\end{align}
\end{subequations}
where the superscript $i$ denotes values at the entrance.

Most simulations set the scale momentum $\po$ such that $\po/q = B_1/h$,
where $B_1$ denotes the design value of the magnetic field strength.
With this choice, the parameter $\eta$ becomes
\[
  \eta = \frac{b_1/h}{B_1/h}
       = \frac{B_1 + \Delta B}{B_1}
       = 1 + \frac{\Delta B}{B_1}
       \equiv 1 + \delta_B,
\]
where $\delta_B$ denotes the relative magnet mispowering error.
For this choice of scaling, and for a correctly powered magnet,
{\ie}~$\delta_B=0$ and $\eta=1$, an on-energy, on-axis particle
will enter the magnet with $\V{z}=0$, and will also exit the
magnet with $\V{z}=0$. In other words, this choice of scaling
means that the map \refE{sbend} with $\eta=1$ preserves the origin.

\subsection{Rectangular bend}
\label{sec:rbend}

For an ideal rectangular bend, one may write the vector potential
term in \refE{generalH} as
\begin{equation}
  -\frac{q}{\po} A_s(X,Y) = \frac{b_1}{\brho}X,
\end{equation}
where $b_1$ and $\brho$ have the same meanings as for the sector
bend. One can again solve the corresponding equations of motion
analytically%
  ~\cite{Forest:1998:BeamDyn}.
Here we define the length parameter
\begin{equation}\label{eq:rhob}
  \rho_b = \frac{\brho}{b_1},
\end{equation}
and we retain the definition of $P_\alpha$ as in \refE{planarP}.
Then, again setting $s=0$ at the magnet entrance,
\begin{subequations}\label{eq:rbend}
\allowdisplaybreaks
\begin{align}
  X(s)   &= X^i + \rho_b\Bigl[
              \sqrt{P_\alpha^2 - \px^2(s)} -
              \sqrt{P_\alpha^2 - (\px^i)^2}
            \Bigr],\\
  \px(s) &= \px^i - \frac{s}{\rho_b},\\
  Y(s)   &= Y^i + \py \,\rho_b
               \left[\sin^{-1}\!\pfrac{\px^i}{P_\alpha}
                    - \sin^{-1}\!\pfrac{\px(s)}{P_\alpha}\right],\\
  \py(s) &= \py^i,\\
  T(s)   &= T^i - \biggl(\frac{1}{\beta_0} + \pt\biggr)
                  \rho_b\left[\sin^{-1}\!\pfrac{\px^i}{P_\alpha}
                             - \sin^{-1}\!\pfrac{\px(s)}{P_\alpha}\right],\\
  \pt(s) &= \pt^i.
\end{align}
\end{subequations}
Remember that, per our discussion near the end of \refS{orbitModel},
the phase-space co\"ordinates here are \emph{internal} to the rectangular
bend. We require transformations that connect these co\"ordinates to
those used outside this magnet.
See \refE{Hprot} and its solution \refE{prot} below.

Since a beam usually enters and exits a rectangular bend at some
non-zero angle $\phi$ with respect to the magnet face, one must
(a) perform dynamic rotations that transform the beam into and
out of the Cartesian frame of the magnet, and (b) account for the
vertical focusing effect. We describe the latter first.

At the entrance, or exit, of a rectangular bend, a typical beam
has a significant horizontal momentum $\px$ with respect to a Cartesian
frame aligned with the corresponding magnet face---approximately the
sine of half the total bend angle of that magnet. Because the magnet
fringe field includes a small horizontal component proportional
(at lowest order) to the vertical displacement, a particle in that
beam experiences a focusing kick (see \refE{vertfocus} below)
\[
  \Delta\py \propto -(\pm\frakb) \px Y,
\]
where $\frakb = eb_1/\po = b_1 / \brho = 1 / \rho_b$ denotes the
scaled dipole field strength, and the upper (lower) sign refers to
the magnet entrance (exit). This kick, however, cannot represent
the whole story, for this by itself cannot constitute a canonical
transformation.

Treating the magnetic field of a rectangular bend as mid-plane
symmetric and as having no dependence on the horizontal co\"ordinate
$X$, one may write the corresponding vector potential in the form
\begin{equation}\label{eq:vecArb}
  \V{A} = \left[\,
            \sum_{k=1}^\infty
                (-1)^k B_\text{o}^{(2k-1)}(z) \frac{y^{2k}}{(2k)!},
            \, 0,\,
            -x B_\text{o}(z)
          \right],
\end{equation}
where $B_\text{o}(z)$ denotes the mid-plane magnetic field. In the
thin-fringe limit, $B_\text{o}(z)$ becomes a step function, and the
$B_\text{o}^{(j)}(z)$ become a delta function and multiple derivatives
thereof. From here there remains a long journey to reach the goal
of a symplectic transfer map that describes orbital motion across
a thin fringe. We simply quote the result.
In the absence of a finite-gap correction, one may obtain%
  ~\cite{Forest:1998:BeamDyn}
\begin{subequations}\label{eq:dipolefringe}
\allowdisplaybreaks
\begin{align}
  X^f   &= X^i + \frac{\pm\frakb}{2\ps}\left\{
             \frac{1 + (x')^2}{1 + (y')^2} -
             \frac{2(x'y')^2}{[1 + (y')^2]^2}
           \right\}\bigl(Y^f\bigr)^2,\\
  \px^f &= \px^i,\\
  Y^f   &= 2Y^i \scalebox{1.5}{/} \left[1
             + \sqrt{1 \pm \frakb
                       \frac{x'y'}{1 + \smash[b]{(y')^2}} \, 2Y^i}\,
             \right]\!,\\
  \label{eq:vertfocus}
  \py^f &= \py^i - (\pm\frakb) \frac{x'}{1+(y')^2} Y^f,\\
  \begin{split}
  T^f   &= T^i - (\pm\frakb)\biggl(\frac{1}{\bo} + \pt\biggr)
           \frac{x'}{2\ps^2}
        \frac{1 - (y')^2}{[1 + (y')^2]^2}\bigl(Y^f\bigr)^2,
  \end{split}\\
  \pt^f &= \pt^i.
\end{align}
\end{subequations}
Here, $x'$ and $y'$ respectively denote $\px/\ps$ and $\py/\ps$,
and $\ps$ remains as given in \refE{ps}. For the result including
a finite-gap correction, consult%
  ~\cite{Forest:2006:FringeEffects1,Forest:2006:FringeEffects2}.

The dynamic rotations that transform the beam into and out of the
Cartesian frame of the magnet amount to drifts in cylindrical
co\"ordinates. Because these transformations rotate the spin, they
are important; on the other hand, they play no r\^ole in the
\emph{accuracy} of our spin integration. Nevertheless, for the sake
of completeness, we include them here: The relevant Hamiltonian,
\begin{equation}\label{eq:Hprot}
  H_\text{prot}
    = -X \sqrt{1 + \frac{2}{\smash{\bo}}\pt + \pt^2 - \px^2 - \py^2}
    = -X \ps,
\end{equation}
yields the solution
\begin{subequations}\label{eq:prot}
\allowdisplaybreaks
\begin{align}
  X(\phi)   &= \frac{X^i \sec\phi}%
                    {1 - (\px^i/\ps^i)\tan\phi},\\
  \px(\phi) &= \px^i \cos\phi + \ps^i \sin\phi,\\
  Y(\phi)   &= Y^i + \frac{\py^i}{\ps^i} \,
                 \frac{X^i\tan\phi}%
                      {1 - (\px^i/\ps^i)\tan\phi},\\
  \py(\phi) &= \py^i,\\
  T(\phi)   &= T^i + \frac{1/\bo + \pt}{\ps^i} \,
                 \frac{X^i\tan\phi}%
                      {1 - (\px^i/\ps^i)\tan\phi},\\
  \pt(\phi) &= \pt^i.
\end{align}
\end{subequations}
One needs to apply this transformation at the entrance with angle
$\phi^\text{entr}$, and at the exit with angle $-\phi^\text{exit}$.
In addition, if the design orbit is not symmetric across the magnet,
\ie~if $\phi^\text{exit}\neq-\phi^\text{entr}$, one must correct for
the horizontal offset
\begin{equation}\label{eq:Xoffset}
  X^\text{offset} = -L\, \frac{\cos\phi^\text{entr} - \cos\phi^\text{exit}}%
                            {\sin\phi^\text{entr} - \sin\phi^\text{exit}}.
\end{equation}
Lastly, if the temporal variable represents a deviation, then one
must subtract from $T$ the amount
\begin{equation}\label{eq:Toffset}
  T_o = -\frac{L_\text{arc}}{\bo}
      = -\frac{L}{\bo} \,
         \frac{\phi^\text{entr} - \phi^\text{exit}}%
              {\sin\phi^\text{entr} - \sin\phi^\text{exit}}
\end{equation}
corresponding to the time for the reference particle to cross this
magnet.


\subsection{Quadrupole}
\label{sec:quad}

For an ideal quadrupole, we write the vector potential term in
\refE{generalH} as
\begin{equation}
  -\frac{q}{\po} A_s(X,Y) = \frac{b_2}{2\brho}(X^2-Y^2),
\end{equation}
where $b_2$ denotes the quadrupole gradient. The most accurate known
split of the resulting Hamiltonian separates out the linear transverse
motion, retaining the exact dependence on the energy deviation~\cite{
Forest:1998:BeamDyn}; thus,
\begin{subequations}\label{eq:MKsplitQuad}
\begin{equation}\label{eq:MKsplitQuadH}
  H_q = H_{q\text{L}} + H_{q\text{NL}},
\end{equation}
where
\begin{equation}\label{eq:mkQuadL}
  H_{q\text{L}} =
    \frac{1}{2}\frac{\px^2 + \py^2}%
                    {\sqrt{1 + \frac{2}{\smash{\bo}}\pt + \pt^2}}
      + \frac{b_2}{2\brho}(X^2 - Y^2),
\end{equation}
and
\begin{equation}\label{eq:mkQuadNL}
  H_{q\text{NL}} =
    - \sqrt{1 + \frac{2}{\smash{\bo}}\pt + \pt^2 - \px^2 - \py^2}
    - \frac{1}{2} \frac{\px^2 + \py^2}%
                       {\sqrt{1 + \frac{2}{\smash{\bo}}\pt + \pt^2}}
    + \frac{P_t}{\bo}.
\end{equation}
\end{subequations}
This is the splitting we use for the quadrupole Hamiltonian.

An alternative to the above split~\cite{Forest:1994:CorrectLocal}
does not include the correction for the energy deviation---the
square root in \refE{mkQuadL}. While that split fails to produce
the correct tune for off-energy particles, it does have the virtue
of speed, because the same transfer matrix applies to all particles.
 
To solve the Hamiltonians in \refE{MKsplitQuad}, we first define
the scaled total momentum
\begin{equation}\label{eq:quadSclP}
  P = \sqrt{1 + \frac{2}{\smash{\bo}}\pt + \pt^2},
\end{equation}
which is a constant of the motion in the quadrupole, and also the
(energy-dependent) focusing strength
\begin{equation}\label{eq:kappa2}
  \kappa = \sqrt{\frac{b_2}{\smash{\brho}\,P}}.
\end{equation}
Then, setting $s=0$ at the magnet entrance, one obtains the solution
for the linear transverse motion, exact in the energy deviation, as 
\begin{subequations}\label{eq:quadL}
\allowdisplaybreaks
\begin{align}
  X\sL(s)   &= X^i \cos(\kappa s)
                 + s \frac{\px^i}{P}\frac{\sin(\kappa s)}{\kappa s},\\
  \px\sL(s) &= \px^i \cos(\kappa s) - \kappa P X^i \sin(\kappa s),\\
  Y\sL(s)   &= Y^i \cosh(\kappa s)
                 + s \frac{\py^i}{P}\frac{\sinh(\kappa s)}{\kappa s},\\
  \py\sL(s) &= \py^i \cosh(\kappa s) + \kappa P Y^i \sinh(\kappa s),\\
  \begin{split}
  T\sL(s)   &= T^i -\frac{1}{P}\left(\frac{1}{\beta_0} + \pt\right)
        \left\{
        \frac{s}{4}\biggl[\frac{(\px^i)^2 + (\py^i)^2}{P^2}
                          + \kappa^2\left((X^i)^2 - (Y^i)^2\right)\biggr]
        \right.\\
    & \qquad\qquad
      + \frac{s}{4}\biggl[\pfrac{\px^i}{P}^2 - (\kappa X^i)^2\biggr]
        \frac{\sin(2\kappa s)}{2\kappa s}\\
    & \qquad\qquad
      + \frac{s}{4}\biggl[\pfrac{\py^i}{P}^2 + (\kappa Y^i)^2\biggr]
        \frac{\sinh(2\kappa s)}{2\kappa s}\\
    & \left.\qquad\qquad- \frac{\kappa s}{2}
        \biggl[X^i \frac{\px^i}{P} \frac{\sin^2(\kappa s)}{\kappa s}
             - Y^i \frac{\py^i}{P} \frac{\sinh^2(\kappa s)}{\kappa s}
        \biggr]\right\},
  \end{split}\\
  \pt\sL(s) &= \pt^i.
\end{align}
\end{subequations}
This solution assumes a positive value for the quadrupole gradient
$b_2$, so that this magnet focuses in the $X$-$\px$ plane and
defocuses in the $Y$-$\py$ plane. We can always convert a skew
quadrupole, or a more general field orientation, to this case by
applying rotations in the transverse plane---\ie~about the longitudinal
axis---before and after the element.

The nonlinear part of the quadrupole Hamiltonian, $H_{q\text{NL}}$,
represents a ``kick'' of the position co\"ordinates. A straightforward
integration yields the solution
\begin{subequations}\label{eq:quadNL}
\allowdisplaybreaks
\begin{align}
  X\sNL(s)   &= X^i + s \px^i
                      \left(\frac{1}{\ps^i} - \frac{1}{P}\right),\\
  \px\sNL(s) &= \px^i,\\
  Y\sNL(s)   &= Y^i + s \py^i
                      \left(\frac{1}{\ps^i} - \frac{1}{P}\right),\\
  \py\sNL(s) &= \py^i,\\
  T\sNL(s)   &= T^i - s\left(\frac{1}{\bo} + \pt\right)
      \left(\frac{1}{\ps^i} - \frac{1}{P}
              \frac{(\px^i)^2 + (\py^i)^2}{2P^2}\right)
       + \frac{s}{\bo},\\
  \pt\sNL(s) &= \pt^i,
\end{align}
\end{subequations}
where
\begin{equation}
  \ps^i = \sqrt{\smash[b]{1 + 2\pt/\bo + \pt^2 - (\px^i)^2 - (\py^i)^2}}.
\end{equation}

In addition to producing the correct tunes for off-energy particles,
the Hamiltonian split shown in \refE{MKsplitQuad} has the additional
advantage that the term $H_{q\text{NL}}$ is small in the paraxial
approximation. In particular, it is fourth-order in the transverse
momenta: $H_{q\text{NL}} \sim \mathcal{O}(\px^m \py^n)$ with
$m+n=4$.

We employ the linear and nonlinear solutions shown here by concatenating
the corresponding maps in a time-reversal-symmetric manner. Using the
simplest symmetric approximation,
\begin{equation}\label{eq:quadConcat}
  e^{s\lieop{H_q}} \approx
      e^{s/2\lieop{H_{q\text{L}}}}
      e^{s\lieop{H_{q\text{NL}}}}
      e^{s/2\lieop{H_{q\text{L}}}},
\end{equation}
we obtain results that are accurate through to second order in the
step size $s$. (Here we have used the colon notation introduced by
Dragt%
  ~\cite{Dragt:1982:NLOrbitDyn,%
         Dragt:1983:CompNLHam}.)
If desired, one may use higher-order versions of
\refE{quadConcat}%
  ~\cite{Yoshida:1990:ConstrHiOrder,%
         McLachlan:2002:SplitMeth,%
         Forest:2006:GeomIntegPA}.

For information about transfer maps for quadrupole fringe fields, consult%
  ~\cite{Forest:1988:HardEdgeFringe}.

\subsection{Higher-order multipole}
\label{sec:mpole}

For an ideal straight multipole magnet---sextupole, octupole,
\etc---one may write the vector potential term in \refE{generalH}
as
\begin{equation}\label{eq:Amult}
  -\frac{q}{\po} A^{(m)}_s(X,Y)
     = \re{\frac{b_m + \ui a_m}{m\brho} (X + \ui Y)^m},
\end{equation}
where the $b_m$ and $a_m$ respectively denote the normal and skew
multipole coefficient for a magnet with $2m$~poles.
We split the resulting Hamiltonian into two pieces:

\begin{subequations}\label{eq:DKsplitMult}
\begin{equation}\label{eq:DKsplitMultH}
  H_\text{mult} = H_\text{drift} + H_m,
\end{equation}
where
\begin{equation}\label{HmultD}
\begin{split}
  H_\text{drift}
    &= - \sqrt{1 + \frac{2}{\smash{\bo}}\pt + \pt^2 - \px^2 - \py^2}
       + \frac{1}{\bo}\pt\\
    &= - \ps + \frac{1}{\bo}\pt,
\end{split}
\end{equation}
\end{subequations}
and $H_m$ denotes the multipole term in \refE{Amult}, or possibly
a superposition of such terms. This constitutes the well-known
\emph{drift-kick} split. Because $H_\text{drift}$ depends only on
momenta, and $H_m$ only on co\"ordinates, they each generate very
simple motion. We then solve the total Hamiltonian, $H_\text{mult}$,
in a manner exactly analogous to our solution in \refE{quadConcat}
of the quadrupole Hamiltonian \refE{MKsplitQuad}.

For information about transfer maps for multipole fringe fields, consult%
  ~\cite{Forest:1988:HardEdgeFringe}.

\subsection{Thin lens}
\label{sec:thinlens}

A thin lens is a representation of a (short) multipole magnet
wherein we imagine the length shrinks to zero and the strengths
$b_m$ and $a_m$ grow to infinity in such a manner that their
product remains constant. Except for taking this limit, we
treat the thin lens just as we do the multipole of \refS{mpole}.
Since the drift vanishes, one may think of this as ``a drift-kick
split without the drift''.

\subsection{TEAPOT}
\label{sec:teapot}

The orbital integration performed in \teapot%
  ~\cite{Schachinger:1985:TEAPOT}
is essentially identical in spirit to the integration of multipoles
described above in \refS{mpole}: One splits the general Hamiltonian
\refE{generalH}
into a drift (in Frenet-Serret co\"ordinates for a sector bend)
and a momentum kick due to the magnetic field. Thus
\begin{subequations}\label{eq:DKsplit}
\begin{equation}\label{eq:HT}
  H_T = H_D + H_K,
\end{equation}
where, in general,
\begin{equation}\label{eq:teapotHD}
  H_D = - (1 + hX)\sqrt{1 + \frac{2}{\smash{\bo}}\pt + \pt^2
                          - \px^2 - \py^2},
\end{equation}
and
\begin{equation}\label{eq:teapotHK}
  H_K = - \frac{q}{p_0} (1 + hX) A_s(X,Y).
\end{equation}
\end{subequations}
When $h$ vanishes, one obtains exactly the multipole drift-kick
split of \refS{mpole}. In the case of non-zero curvature, \ie\ $h\neq0$,
the vector potential $A_s$ becomes quite complicated. Nevertheless,
because $H_K$ depends only on co\"ordinates, one may easily compute
the motion generated.

The drift Hamiltonian $H_D$ differs only slightly from that in
\refE{Hprot}, and we obtain the solution from \refE{prot} by replacing
$X$ with $1/h + X$, and $\phi$ with $hs$:
\begin{subequations}\label{eq:teapotD}
\allowdisplaybreaks
\begin{align}
  X(s)   &= \frac{(1/h + X^i)\sec(hs)}%
                 {1 - (\px^i/\ps^i)\tan(hs)}
            - \frac{1}{h},\\
  \px(s) &= \px^i \cos(hs) + \ps^i \sin(hs),\\
  Y(s)   &= Y^i + \frac{\py^i}{\ps^i} \,
              \frac{(1/h + X^i)\tan(hs)}%
                   {1 - (\px^i/\ps^i)\tan(hs)},\\
  \py(s) &= \py^i,\\
  T(s)   &= T^i - \frac{1/\bo + \pt}{\ps^i} \,
              \frac{(1/h + X^i)\tan(hs)}%
                   {1 - (\px^i/\ps^i)\tan(hs)},\\
  \pt(\phi) &= \pt^i.
\end{align}
\end{subequations}
As in \refE{quadConcat}, we then combine the two solutions---the
one above for $H_D$, and the one for $H_K$---to approximate the
motion generated by the full Hamiltonian $H_T$.

A disadvantage of this integration scheme is that the two pieces
of the Hamiltonian $H_T$ in \refE{HT} may be of similar order.
While the resulting integrator is still second-order accurate in
$s$, it has a larger constant factor in the error term. This yields,
overall, a less accurate integration for a given number of slices.
This proves to be a significant issue when we go on to integrate
the spin motion.

On the other hand, drift-kick integration has the advantage that it
applies to any magnetic field described by a longitudinal vector
potential  $A_s(X,Y)$. Moreover, it does not suffer from the apparent
discontinuity in the map that occurs for the quadrupole when using
the matrix-kick integrator of \refS{quad}. Avoiding the use of
control-flow statements [\texttt{if(focusingQuad) \{\dots\}}
\texttt{else} \{\dots\}] can be an
important consideration when using an integrator for map computation
rather than particle tracking%
  ~\cite{Forest:2006:GeomIntegPA}.

\subsection{Solenoid}
\label{sec:solenoid}

For the solenoid, we use a model which has a uniform longitudinal
field in the body and the flux return confined to a thin-pancake
fringe region. The longitudinal body field necessitates a transverse
vector potential $(b_0/2)\rho\hat\phi$, which we write in the Cartesian
form
\begin{equation}\label{eq:Asol}
  A_x = -\frac{1}{2} b_0 Y,
  \quad
  A_y =  \frac{1}{2} b_0 X,
\end{equation}
where $b_0$ denotes the magnetic field strength.

For our thin-pancake fringe model, we multiply the components
\refE{Asol} by a profile function $S_\varepsilon(z)$ that rises
from zero to one across an entrance fringe of length $\varepsilon$,
remains constant across the body, and then returns to zero across
an exit fringe of length $\varepsilon$. We eventually take the
limit $\varepsilon\rightarrow0$. This model means our pancake
fringe has a longitudinal field that rises, or falls, linearly
across the fringe, and a radial field with flux $b_0\pi\rho^2$
crossing an area $2\pi\rho\,\varepsilon$ at radius $\rho$.

Because we have a transverse vector potential, we must modify
the general Hamiltonian \refE{generalH}: Since we use a Cartesian
reference frame, we set $h=0$; of course we set $A_s=0$; and the
canonical momentum now equals kinetic momentum plus vector potential.
We therefore write the Hamiltonian for a magnetic solenoid
in the form
\begin{equation}\label{eq:Hsolenoid}
\begin{split}
  H_\text{sol}
    &= -\left[1 + \frac{2}{\bo}\pt + \pt^2
                - \Bigl(\px - \frac{e}{\po}A_x\Bigr)^2
                - \Bigl(\py - \frac{e}{\po}A_y\Bigr)^2
            \right]^{1/2} + \frac{1}{\bo}\pt \\
    &= -\ps + \frac{1}{\bo}\pt.
\end{split}
\end{equation}

On crossing one of the solenoid fringes, a particle experiences
a transverse kick from the radial component of the magnetic
field. The transverse vector potential also changes across the same
fringe. It turns out that for our thin-pancake fringe model, those
changes---to the kinetic momentum and the vector potential---cancel
in a manner that leaves the transverse canonical momentum unchanged
across the fringe.
Moreover, 
in the small-$\varepsilon$ limit, the transverse co\"ordinates do not
change across a fringe. Finally, we ignore the effect on the
longitudinal momentum, because it is second-order in the
transverse dynamical variables. The net result is that
for our thin-pancake fringe, the Hamiltonian \refE{Hsolenoid}
with vector potential \refE{Asol} describes the entire solenoid.

Given the above model, one can compute an exact analytic result
for the motion generated by the Hamiltonian \refE{Hsolenoid}.
The solution comprises the product of a pair of commuting matrices
that act on the transverse variables $(X,\px, Y,\py)$, plus a shift
of the temporal variable $T$:
\begin{subequations}\label{eq:solenoid}
\begin{equation}\label{eq:rotPS}
R_1 = \left[
\begin{array}{cccc}
  \cos(\kappa s)
    & \frac{1}{\eta}\sin(\kappa s)
    & 0 & 0 \\[1.2ex]
  -\eta\sin(\kappa s)
    & \cos(\kappa s)
    & 0 & 0 \\
  0 & 0
    & \cos(\kappa s)
    & \frac{1}{\eta}\sin(\kappa s) \\[1.2ex]
  0 & 0
    & -\eta\sin(\kappa s)
    & \cos(\kappa s)
\end{array}
\right]
\end{equation}
\begin{equation}\label{eq:rotXY}
R_2 = \left[
\begin{array}{cccc}
  \cos(\kappa s)
    & 0
    & \sin(\kappa s)
    & 0 \\[0.7ex]
  0 & \cos(\kappa s)
    & 0
    & \sin(\kappa s) \\[0.7ex]
  -\sin(\kappa s)
    & 0
    & \cos(\kappa s)
    & 0 \\[0.7ex]
  0 & -\sin(\kappa s)
    & 0
    & \cos(\kappa s)
\end{array}
\right]
\end{equation}
\begin{equation}\label{eq:solT}
  T(s) = T^i - \Bigl(\frac{1}{\bo} + \pt\Bigr)\frac{s}{\ps^i}
             + \frac{s}{\bo}.
\end{equation}
\end{subequations}
In the above expressions, the parameter
\begin{equation}
  \eta = \frac{e b_o}{2\po} = \frac{b_o}{2\brho},
\end{equation}
$\ps^i$ denotes the initial value of the square-root in the
Hamiltonian \refE{Hsolenoid}, and $\kappa=\eta/\ps^i$.

The reader should note that the presence of $\ps^i$ in
\refE{solenoid}  means that---despite superficial
appearances---the transformation represented by \refE{solenoid}
is actually nonlinear.

Later, when computing the spin precession vector $\bigOmega$,
we shall need to compute the direction vector $\uhat$. All
magnets discussed previously can be described using a longitudinal
vector potential. Then, as mentioned in \refS{spinModel}, we
can obtain $\uhat$ simply by normalising the momentum vector
$(\px,\py,\ps)^T$. For the solenoid, however, we must be sure
to subtract the transverse components of the vector potential
from the transverse canonical momenta to obtain the (scaled)
kinetic momentum vector
\begin{equation}\label{eq:mgVSol}
  m\gamma\V{V} = \Bigl(
    \px - \frac{e}{\po}A_x,\,
    \py - \frac{e}{\po}A_y,\,
    \ps \Bigr)^T.
\end{equation}
From this we compute $\uhat = m\gamma\V{V} / \|m\gamma\V{V}\|$.

\section{Spin Integration}
\label{sec:spinInteg}

High-quality orbital integration is a prerequisite for accurate
spin integration, because it yields accurate values for the
precession vector $\bigOmega$. However, discretizing the spin
motion leads to a separate source of error that can cause
inaccurate spin precession, even in the context of perfect orbital
integration. To address this difficulty, one might, for example,
choose to adjust the number of spin slices independently of the
number of orbital slices. In this section, we discuss this
issue along with a description of our methods for tracking
spin.

\subsection{Piece-wise constant (\pwc) spin precession}
\label{sec:spinIntegPWC}

A commonly used integration strategy for spin treats the
magnetic fields and velocity vectors in \refE{t-bmt.so} as
constant throughout a slice of length $\Delta s$. One then
integrates \refE{t-bmt.s}, the \tbmt\ equation, to the form
\begin{subequations}\label{eq:pwcSpin}
\begin{equation}\label{eq:spinSlice}
  \spin(s+\Delta s) = \mathbf{R}(\V\omega) \cdot \spin(s),
\end{equation}
where $\mathbf{R}(\V\omega)$ denotes the $3\times3$ matrix that
describes rotation about axis $\V\omega$ by angle $|\V\omega|$.
We approximate this rotation vector as $\V\omega=\Delta{s}\,\bigOmega$,
with $\bigOmega$ given in \refE{t-bmt.so}. Then to compute the spin
rotation across a whole element, one simply multiplies the
contributions from each slice. For four slices, say, one thus
transports an initial spin $\spin^i$ to a final spin $\spin^f$
according to
\begin{equation}\label{eq:si2sf}
  \spin^f = \mathbf{R}(\V\omega_4) \cdot
            \mathbf{R}(\V\omega_3) \cdot
            \mathbf{R}(\V\omega_2) \cdot
            \mathbf{R}(\V\omega_1) \cdot \spin^i.
\end{equation}
\end{subequations}

There are two sources of errors in this approach to spin integration.
First, errors in the orbital integration feed into the spin
integration via errors in the fields and velocities needed in the
right-hand side of the \tbmt\ equation. Second, treating the rotation
axis as piece-wise constant introduces errors that arise from the
non-commutativity of spin rotations around non-parallel axes.

When using drift-kick integrators for the orbital motion, the first
source of error usually dominates. Then increasing the number of
orbital slices to improve the orbital accuracy automatically diminishes
the magnitude of the second type of error. We thus find that with
drift-kick integrators, increasing the number of orbital slices
is essential for accurate spin integration.

The situation changes when we use the more accurate bend-kick and
matrix-kick integrators described above for dipoles and quadrupoles.
These integrators allow us to take such large steps through both
dipoles and quadrupoles that the lack of commutativity between
consecutive spin rotations can now become an issue.

To speed the accumulation of spin rotations across a set of orbital
slices, we use quaternions to represent rotations%
  ~\cite{Biedenharn:1981:AngMomQP}.
Compared to matrix multiplication, this saves a factor of about two
in the required arithmetic operations. In addition, the spin rotation
angles are very often sufficiently small that one may approximate
the needed trigonometric functions by low-order polynomials. Every
so often, we explicitly normalize the resulting quaternions to
ensure that they remain unit quaternions.%
  \footnote{To simplify the bookkeeping, we do this at the end of
            each element, but once per turn would likely suffice.}

\subsection{B\'ezier interpolation}
\label{sec:bezier}

\begin{figure}
  \includegraphics{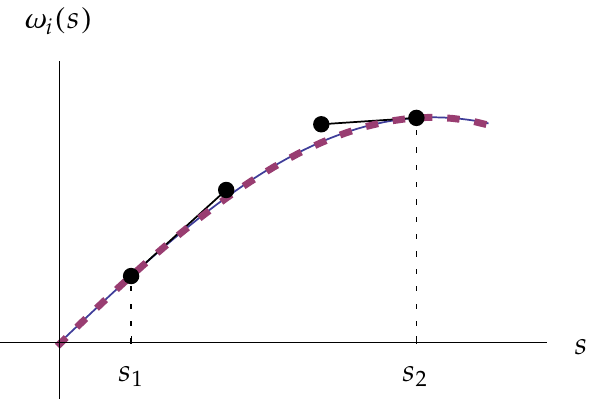}
  \caption{Schematic of interpolating a function using a cubic
B\'ezier curve. The function to be interpolated is shown as a thin
solid curve,
and the constructed interpolating B\'ezier curve is shown by the
heavy dashed curve. The large dots indicate the control points
derived from the value and slope of the function at $s_1$ and $s_2$.}
  \label{fig:cubicBezier}
\end{figure}

To treat the second source of error---arising from the fact that
non-parallel rotations do not commute---we could do the same as
for the first source of error: increase the number of orbital
slices. That approach, however, seems unnecessarily expensive
when we use very accurate, or exact, orbital integrators.

It turns out that the rotation vector $\bigOmega$ varies across
a slice in a sufficiently simple manner that we may interpolate
it across a slice using piece-wise cubic polynomials, constructed
as B{\'e}zier curves%
  ~\cite{Casselman:2005:MathIllustr}.
To use this technique, we compute, at the edge of each slice
during the orbital integration, not only $\bigOmega$, but also
the derivatives of its three components.%
  \footnote{%
Computing the longitudinal derivative of $\bigOmega$ is tedious,
but straightforward: Express the magnetic field in terms of particle
location---so that $\bigOmega$ depends solely on the orbital
phase-space variables. Then use the chain rule to compute
\begin{equation*}
  \Dd{\bigOmega}{s} = \ptDd{\bigOmega}{X} \ptDd{X}{s}
                     +\ptDd{\bigOmega}{\px} \ptDd{\px}{s}
                     +\dotsb.
\end{equation*}
The quantities $\partial_s X$, $\partial_s \px$, \ldots, can
be obtained from the orbital equations of motion.
}
With that information,
we construct a cubic interpolating polynomial with the correct
slope at the end points; see \refF{cubicBezier} for a schematic
illustration. For the $\bigOmega$ variations that occur in our
simulations, this approach yields very accurate interpolating
functions. Moreover, the low cost of evaluating this interpolating
polynomial makes it possible to take many spin steps per orbital
slice.

The \pwc\ computation of the spin precession in \refE{si2sf}
will converge with a sufficiently large number of steps. If
convergence requires an excessive number of steps, the use of
cubic B\'ezier curves can speed up the computation. What the
B\'ezier approach does \emph{not} do is increase the rate at
which the computation converges. For that we must look elsewhere,
and that is the topic of the next section.

\subsection{Romberg quadratures for spin precession}
\label{sec:RombergSpin}

As we have seen,
the spin precession across a beamline element is not, in reality,
piece-wise constant; rather, within an element it varies smoothly
with $s$. More significantly, the integrators based on \refE{pwcSpin}
exhibit second-order convergence (see \refS{spinPerform}).
This quadratic convergence suggests the use of some accelerating
technique to cancel the errors. In particular, we have applied a
Romberg approach%
  ~\cite{Acton:1996:RealComp,%
         Press:2002:NumerRecipes}
to spin integration, and, as we show later in this paper, the
improvement is dramatic.

Instead of computing the rotation vectors at the middle of each slice,
we now, for this new approach, compute them at the edges of each
slice. We then accumulate the spin precession as in \refE{pwcSpin}.
For the first and last edges, however, we use a half-step; \ie\
we replace $\Delta s$ by $\Delta s/2$. This is akin to using the
trapezoidal rule for integration%
  ~\cite{Acton:1996:RealComp,%
         Hildebrand:1987:IntroNumerAnal}.
We do this using $N$ slices, with $N$ a multiple of some power of two.

During the orbital integration, we record the value of $\bigOmega$ at
the edge of each slice. Then by keeping every other $\bigOmega$, or
every fourth, or every eighth, \etc, we can approximate the net spin
precession using a range of additional step-sizes that are related
to the original by powers of two. In other words, we compute for
these different step sizes a net quaternion that represents the spin
precession of a given particle across an element. We obtain thus the
sequence of quaternions
\begin{equation}\label{eq:Qhs}
  Q(h),\; Q(h/2),\; Q(h/2^2),\; Q(h/2^3),\; \dotsc,
\end{equation}
where $h$ denotes the coarsest step size.

For example, when crossing an element using eight slices, we will
compute a total of nine $\bigOmega$s: $\bigOmega_0$ at the beginning
of the first slice, $\bigOmega_1$ at the end of the first slice,
on through $\bigOmega_8$ at the end of the last slice. Now, assuming
each slice has length $\Delta s$, we multiply the first and last of
these by $\Delta s/2$, and all the others by $\Delta s$, to obtain
the sequence
\begin{equation*}
  \frac{1}{2}\Delta s\,\bigOmega_0,\ \Delta s\,\bigOmega_1,
  \ \Delta s\,\bigOmega_2,\ \dotsc,\ \frac{1}{2}\Delta s\,\bigOmega_8.
\end{equation*}
After constructing the quaternions that correspond to each of these
nine precession vectors, we multiply them together to obtain a net
quaternion that represents one approximation to the exact spin precession
across this element. This is one of the entries in the sequence \refE{Qhs}
above. To obtain the entry to its immediate left, we drop every other
$\bigOmega$ and compensate by doubling the step size; the sequence
\begin{equation*}
  \Delta s\,\bigOmega_0,\ 2\Delta s\,\bigOmega_2,\ \dotsc,
  \ \Delta s\,\bigOmega_8,
\end{equation*}
thus yields another---the next coarser---approximation to the spin
precession across this element. And so on.

We then compute the \emph{Romberg limit}: First define by
\begin{subequations}\label{eq:romberg}
\begin{equation}\label{eq:Q0k}
  Q_{0k} = Q(h/2^k)
\end{equation}
the sequence of approximations in \refE{Qhs}. Then use the rule
\begin{equation}\label{eq:rombergStep}
  Q_{j+1,k} = \frac{4^{j+1} Q_{j,k} - Q_{j,k-1}}{4^{j+1} - 1}
\end{equation}
to construct the Romberg table:
\begin{equation}\label{eq:rombergTab}
  \begin{array}{ccccccc}
    Q_{00} &&        &&        &&        \\
    Q_{01} && Q_{11} &&        &&        \\
    Q_{02} && Q_{12} && Q_{22} &&        \\
    Q_{03} && Q_{13} && Q_{23} && Q_{33}.
  \end{array}
\end{equation}
\end{subequations}
This table may have more or fewer rows than indicated here,
but the number at the bottom right is the Romberg limit of the
initial data given in the first column. We normalise the
resulting quaternion at the end of each element.

When integrating a well-behaved function over a finite interval,
the trapezoidal rule together with the Romberg limit performs
remarkably well with modest computational effort. Its efficiency,
however, derives from the structure of the error term seen in the
Euler-Maclaurin summation formula%
  ~\cite{Hildebrand:1987:IntroNumerAnal,%
         Press:2002:NumerRecipes},
and the manner in which the Romberg algorithm cancels those errors.
Here, on the other hand, we have a \emph{product} of quaternions,
and hence no \emph{a priori} reason to suspect that the above will
actually work. We tried it for a lark.

\subsection{Solenoid fringe}
\label{sec:solFringe}

In the thin-pancake model discussed earlier for the solenoid fringe,
there exists a well-defined limit of the fringe length $\varepsilon$
times the precession vector $\bigOmega$, as $\varepsilon\rightarrow0$.
In that limit, a spin crossing a solenoid fringe experiences a net
rotation given by the vector
\begin{equation}\label{eq:solFringe}
  \V\omega = \pm\frac{b_0/2}{\brho\ps}
    \Bigl[(1 + G\gamma)(X,\,Y,\,0)
           - G(\gamma - 1)\frac{X\px + Y\py}{P^2} m\gamma\V{V}\Bigr],
\end{equation}
where we choose the leading plus or minus for entrance or exit,
respectively. Here the kinetic momentum vector $m\gamma\V{V}$
is as given in \refE{mgVSol}, \emph{except} that we must insert
a factor of $\frac{1}{2}$ in front of $A_x$ and $A_y$. In a similar
manner, $\ps$ denotes the square-root in \refE{Hsolenoid}, but with
the same factor of $\frac{1}{2}$ inserted before $A_x$ and $A_y$.
These factors of $\frac{1}{2}$ enter because in going to the
limit of a zero-length fringe, we evaluate the field strength in
the middle of that fringe region.

\subsection{Dipole and multipole fringes}
\label{sec:mpFringe}

A particle's spin also experiences a kick when
crossing the fringe of a multipole magnet. This spin kick arises
predominantly from the longitudinal field component in the fringe.
In the limit of a thin fringe, and to lowest order in the
dynamical variables, the spin kick caused by the fringe of a
normal (\ie~not skew) $2m$-pole magnet of strength $b_m$ is
given by the rotation vector
\begin{equation}\label{eq:skfringe}
  \V\omega_m = -\frac{\pm b_m}{\brho}
              (1 + G)\frac{r^m}{m}\sin(m\phi)\,\zhat,
\end{equation}
where we convert $r^m\sin(m\phi)$ to a function of $X$ and $Y$ using
\(
  (X + \ui\mhsp Y)^m = r^m \ue^{\ui m\phi}.
\)
For a skew multipole, in \refE{skfringe} replace $\sin$ by $\cos$,
and $b_m$ by $a_m$.

The above result \refE{skfringe} is appropriate for beams that
cross a fringe roughly normal to the entrance or exit face of the
magnet. When entering, or exiting, a rectangular bend, however,
a particle typically sees the longitudinal field $B_z$ as having
a significant component transverse to its velocity. As long as 
$\px$ is not too large, one may account for this effect by simply
adding an extra term to \refE{skfringe} to obtain the rotation
vector
\begin{equation}\label{eq:skfringeRB}
  \V\omega_\text{rb} = \V\omega_1
    + \frac{\pm b_1}{\brho}
              G(\gamma - 1)\px\,\xhat,
\end{equation}
which describes the spin kick experienced by a particle crossing
the fringe of a rectangular bend. Note the factor of $G\gamma$
in the second term: it is one good reason for not using
rectangular bends in machines designed for polarized beams,
particularly at high energy.

If one requires a higher order computation of the spin kick
across a magnet fringe, it is possible to apply a procedure
analogous to that used for orbital kicks across fringe fields
in%
  ~\cite{Forest:1998:BeamDyn}.

\section{Performance of integrators}
\label{sec:performance}

In this section, we examine the performance of the orbital and
spin integrators described in the previous two sections. Our
principal focus is on the integrators for spin. We examine the
orbital integrators---much more briefly---with the goal of
understanding their impact on spin integration.

For the orbital integrators, we examine their performance in a
particular context: that of Brookhaven's Relativistic Heavy
Ion Collider (\rhic)%
  ~\cite{Harrison:2002:RHICaccelerator}
operating at approximately \SI{250}{\GeV}, with the optics settings used
before the beta squeeze. For the spin integrators, we examine
their single-element performance, as well as their performance in
the context of spin tracking for \rhic. In this context, we note
that \rhic\ has two full Siberian snakes on opposite sides of the
ring to flip the spins, with the snake angles set to $\pm\ang{45}$.
These settings mean that for a perfectly aligned \rhic, the design
orbit will have a spin-tune of exactly $\frac{1}{2}$%
  ~\cite{Bai:2008:RHICStatusPP,Huang:2011:RHICStatus}.
Sextupoles are adjusted to set the chromaticities to
$(\xi_x,\xi_y)=(0.70,0.74)$, and the spin rotators near the
interaction points are turned off.
The latter are needed only after the beta squeeze, when the
counter-rotating \rhic\ beams are brought to collision, and
the experiments require longitudinal polarization at the
interaction point. For our simulations, then, the equilibrium
polarization is roughly vertical throughout the ring.

\subsection{Orbit integrators}
\label{sec:orbitPerform}

For the simulations discussed here, we read a \rhic\ lattice
description from a file in SXF format%
  ~\cite{Grote:1998:SXF}.
Which integrator to use for which elements is set in a separate
`.apdf' file (for \textsl{A}ccelerator \textsl{P}ropagator
\textsl{D}escription \textsl{F}ormat).
When referring to the integrators, we use the acronym \bkmk,%
  \footnote{Bend-Kick/Matrix-Kick}
or sometimes just ``bend-kick'', to mean the use of \refE{sbend}
for sector bends, together with \refE{quadL} and \eqref{eq:quadNL}
for quadrupoles. 
The Siberian snakes were modelled as thin elements: they are
transparent to the orbital motion, and they act on the spin
according to snake angles defined in an input file.
In the simulations described here, we use a base number $n_b$
of slices for all dipoles and most quadrupoles. Then for the
strongest quadrupoles, those in the interaction region, we use
$4n_b$ slices.


\subsubsection{Poincar\'e sections}

\begin{figure}[t]
  \includegraphics{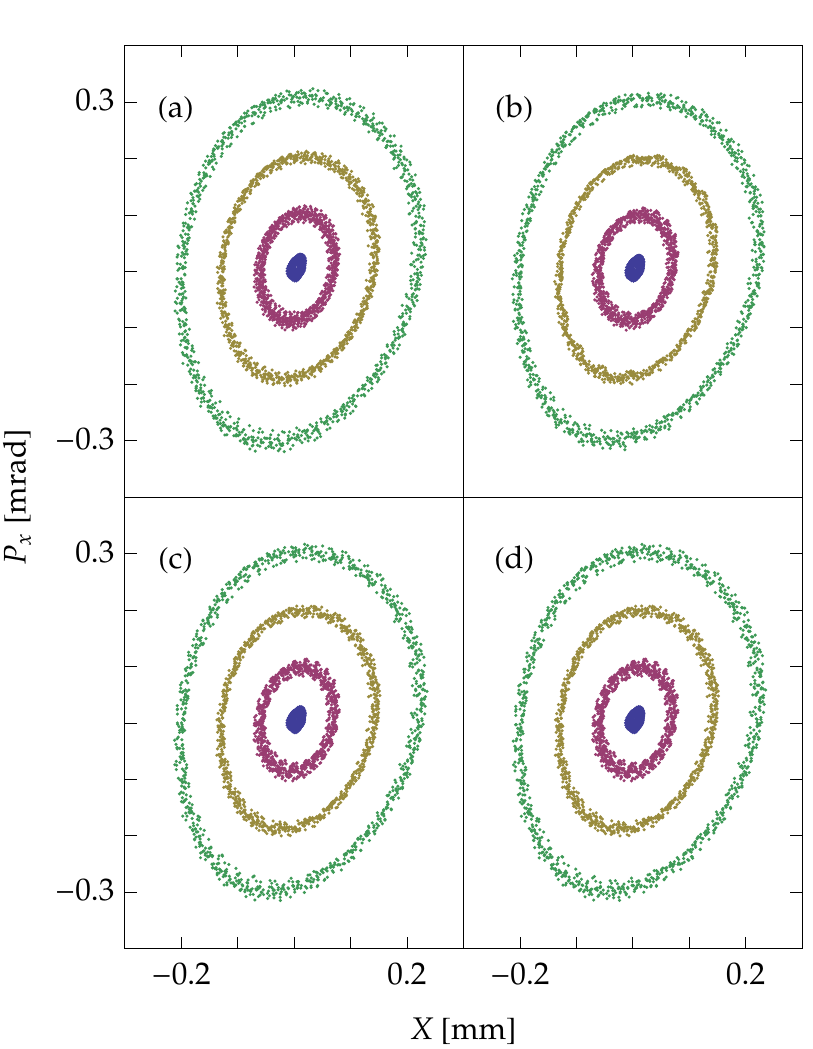}
  \caption{Poincar\'e sections for four sample particles using
drift-kick integration with (a) $n_b = 4$ and (b) $n_b = 16$, and
using \bkmk\ integration with (c) $n_b = 1$ and (d) $n_b = 16$.}
  \label{fig:Poincare}
\end{figure}

To give a sense of how well the drift-kick and \bkmk\ 
integrators reproduce gross features of the orbital motion,
we show Poincar\'e sections of the $X$-$\px$ plane; see
\refF{Poincare}. The range of amplitudes shown is
representative of beams at this azimuth in \rhic---the
interaction point, before the beta squeeze is applied to the
optics.

\refFc{Poincare} indicates that the drift-kick and \bkmk\ 
integrators produce the same gross features of the orbital motion.
There are, however, differences in the details. In particular,
the drift-kick traces with $n_b = 4$ seem more elliptical, while
the remaining traces seem more ovoid. The
trace-widths also seem more regular for drift-kick integration with
$n_b = 4$, as compared to all the others. Differences between the two
\bkmk\ integrators, with $n_b = 1$ and $n_b = 16$, are not visible
with the bare eye. Differences between (b), (c), and (d) are much
smaller than those between (a), on the one hand, and (c) and (d),
on the other.

\subsubsection{Orbital spectra and tune fitting}

\begin{figure}
  \includegraphics{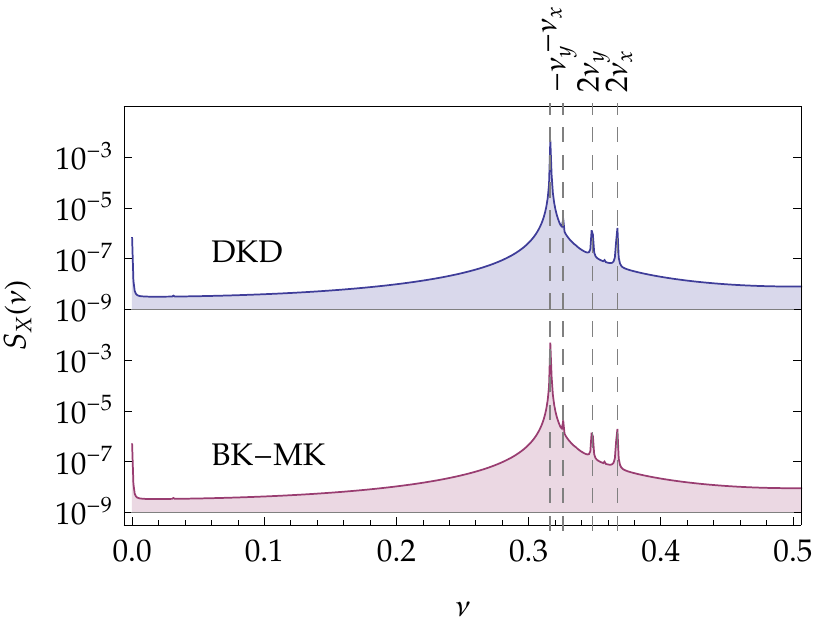}
  \caption{Spectrum of $X$ obtained using (a)~drift-kick integration
with $n_b = 8$, or (b)~\bkmk\ integration with $n_b = 1$.
The indicated tunes are all taken modulo~1.}
  \label{fig:orbitSpectra}
\end{figure}

In \refF{orbitSpectra} we show the spectrum of $X$ obtained
by taking a discrete Fourier transform of the $X$~co\"ordinates
obtained over the course of $T=1000$ turns,
\begin{equation}
  S_{X}(\nu) = \left|\frac{1}{T}\sum_{t=0}^T
                 \ue^{-\ui 2\pi t \nu} X(t)\right|^2.
\end{equation}
The spectra were averaged over a bunch containing $N = 1024$
particles with amplitudes typical for \rhic. The tunes marked
in the figure (see top edge) are the design values for \rhic\ at
this energy (\SI{215.735}{\GeV} kinetic): 
$\nu_x = 28.6835$, $\nu_y = 29.6742$.
Note that we see peaks not only at $\nu_x$ and $\nu_y$, but
also at $2\nu_x$ and $2\nu_y$---evidence of significant
nonlinearity in the orbital motion.

As shown in \refF{orbitSpectra}, one can
obtain a close match to the design orbital tunes using either
drift-kick or \bkmk\ integration. However, in the case of
drift-kick integration, one must adjust the strengths
of the main quadrupoles by an amount that depends on the number
of orbital slices used. No such adjustment is required when
using \bkmk\ integration.

In the context of purely orbital tracking, adjusting the
quadrupole strengths to fit the desired tunes is a sensible means
of ensuring the correct linear behavior in accelerator simulations.
In the context of spin tracking, however, such adjustments
might be problematic, because changing a quadrupole's strength
changes the amount of spin precession a particle experiences as
it crosses that element. In other words, fitting the orbital
tune can perturb the integrated spin precession angle of a given
particle---particularly at high energy, \ie, at large values of
the na\"ive spin tune, $G\gamma$.

We can estimate the importance of this effect as follows.
On crossing a quadrupole, a spin experiences a net precession
with magnitude given approximately by
\(
  \omega = -(1+G\gamma) A b_2 L/\brho,
\)
where $A$ denotes the particle amplitude, and $L$ the quadrupole
length. For the \rhic\ lattices we studied, the tune fitting
required a relative change in $b_2$ that varied greatly with
the step-size used for crossing the quadrupoles. For finely
sliced lattices, the relative change in $b_2$ (from the value
used for \bkmk\ integration) could be as small as a few times
\num{e-5}. At \rhic\ energies, this adjustment yields
a sub-\si{\murad} change in the spin precession, and perhaps a
few \si{\murad}s over a full turn. As we shall see later, when
one uses drift-kick orbital integration, this variation is
negligible compared to other sources of error. On the other hand,
for very coarse slicing, the relative change in $b_2$ can exceed
\SI{0.1}{\percent}, which leads to a small, though noticeable,
change in the simulated spin precession angle.

\subsubsection{Single-turn errors}

\begin{figure}
  \includegraphics{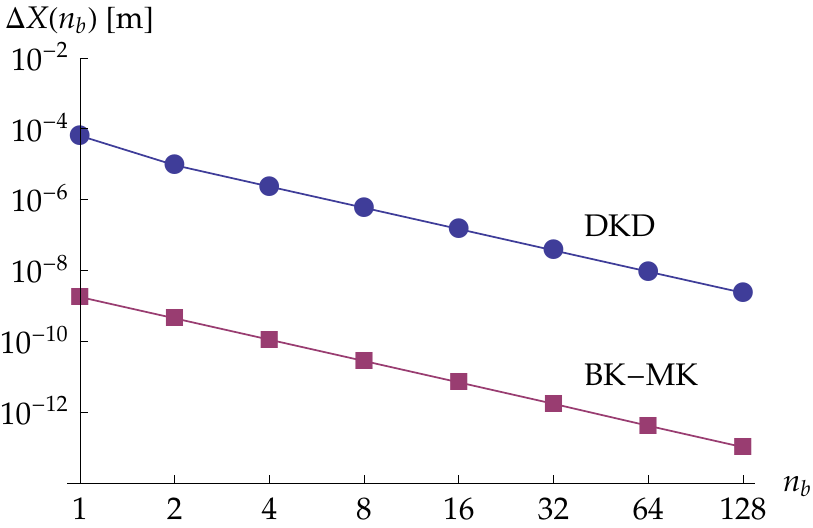}
  \caption{One-turn errors in $X$ versus $n_b$ for drift-kick
integration (upper curve), and \bkmk\ integration (lower curve).
The errors were computed by comparing with a data set produced
using \bkmk\ integration and $n_b = 256$.}
  \label{fig:oneTurnErrsX}
\end{figure}

For accurate spin tracking, it does not suffice to integrate
the orbital motion in a manner that merely reproduces global
properties of the phase-space distribution. Errors in particle
orbits introduce errors in the spin rotations, possibly changing
the character of the simulated spin dynamics. As a consequence,
one must compute accurately the trajectories of individual
particles.

To gain an understanding of the orbital errors on a per-particle
basis, we look at the rms deviation between computed particle
orbits and a reference solution for each particle. For the $X$
co\"ordinate after $t$ turns, we thus compute
\begin{equation}
  \Delta X(t;n_b) = \sqrt{\frac{1}{N} \sum_{\smash{j}=\smash{1}}^N
    \Bigl[X_{j}(t;n_b) - X_{j}^\text{(ref)}(t)\Bigr]^2\,}\,.
\end{equation}
The sum is over the particles in the bunch. The reference solution
is computed with $n_b = 128$ using the \bkmk\ integrators,
\ie~$X_{j}^\text{(ref)}(t) = X_j(t;128)$.

\refFc{oneTurnErrsX} shows the absolute error $\Delta X(1;n_b)$
after one turn for a \rhic\ beam at approximately \SI{216}{\GeV}
with a \SI{95}{\percent} horizontal emittance $\varepsilon_x$ of
\SI{15\pi}{\mm\,\mrad}. The slope \num{-2} on the $\log$-$\log$
scale shows that both drift-kick and \bkmk\ integrators exhibit
second-order convergence in $s$. The drift-kick integrator starts
out with a relative error of order one (\cf~\refF{Poincare}),
indicating that with this very large step-size, the
phases of the orbital motion are completely off: the errors are
as large as they can be with a symplectic integrator. At $n_b = 64$,
the relative errors have decreased to about a part in \num{e4}.
On the other hand, the \bkmk\ integrators exhibit relative
errors which start out below a part in \num{e6} and decrease
until they are below about a part in \num{e9}.

\begin{figure}
  \includegraphics{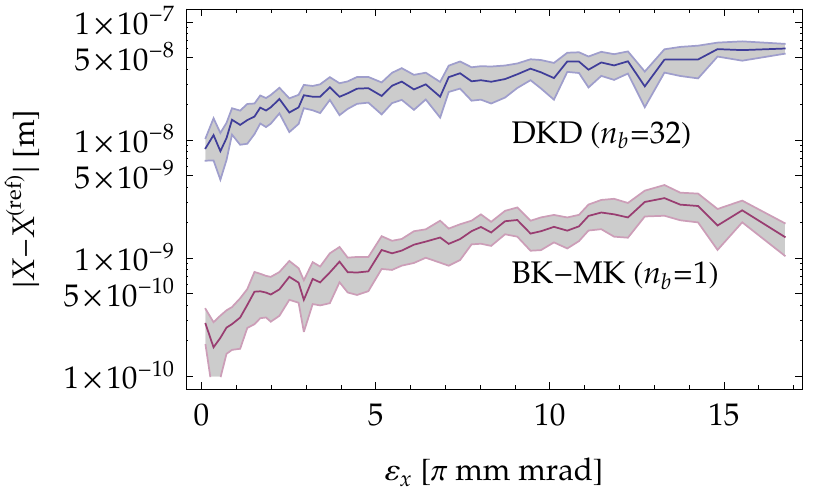}
  \caption{One-turn errors in $X$  versus amplitude for several
different integrators: drift-kick with $n_b = 32$ (upper curve),
and \bkmk\ with $n_b = 1$ (lower curve).}
  \label{fig:xErrVsAmp}
\end{figure}

\refFc{xErrVsAmp} also shows one-turn errors in $X$, but here
as a function of particle emittance. These results, which are
consistent with those of \refF{oneTurnErrsX}, tell us that the
accuracies of the different integrators have roughly the same
amplitude dependence. To reduce the noise level in this graphic,
we have averaged the results within non-overlapping bins that
each contained \num{20} particles in a narrow slice of emittance.
The bands indicate one standard deviation above and below the
computed average.


\subsubsection{Evolution of orbital errors over many turns}

\refFc{xErrVsTurnNo} shows the errors in the $X$ co\"ordinate as
a function of the number of turns for drift-kick integration and
\bkmk\ integration. If we do not adjust the quadrupole strengths
so as to fit the desired tunes, then the drift-kick integration
errors oscillate periodically because of the tune errors. We see
that behavior in the uppermost (tan) curve, for drift-kick integration
with $n_b = 8$. After fitting the tunes, we note that drift-kick
integration errors with $n_b = 4$ now track the minima of the
tan oscillatory errors.

While the errors observed are much smaller for
\bkmk\ integration---even with $n_b = 1$---they always exhibit
the same qualitative behavior: After a few hundred turns, even
when using \bkmk\ integration, the errors grow approximately
linearly because of unavoidable tiny
differences in the tunes. We believe this is not a significant
issue.

\begin{figure}
  \includegraphics{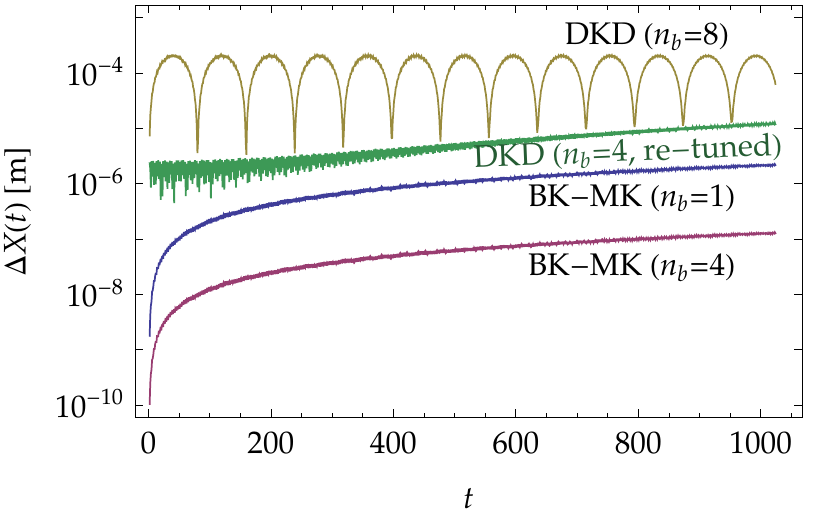}
  \caption{Evolution of the orbital error in $X$ as a function of
turn number $t$. The four curves correspond to different methods
of integration: from top to bottom they are
drift-kick with $n_b = 8$ and no tune fit (tan),
drift-kick after tune fitting with $n_b = 4$ (green), and
\bkmk\ with $n_b = 1$ (purple), and $4$ (maroon).
For the reference solution, we used \bkmk\ with $n_b = 16$.}
  \label{fig:xErrVsTurnNo}
\end{figure}

\subsection{Spin integrators}
\label{sec:spinPerform}

Here we describe the accuracy of our spin integrators. First, we
discuss the errors seen when tracking across various individual beamline
elements, where we focus on the impact of both step-size and Romberg
iterations. We then discuss, more briefly, the spin integration errors
seen over the course of a single turn and many turns.

\enlargethispage{\baselineskip}
\subsubsection{Single-element errors}

\begin{figure}
  \includegraphics[scale=0.825]{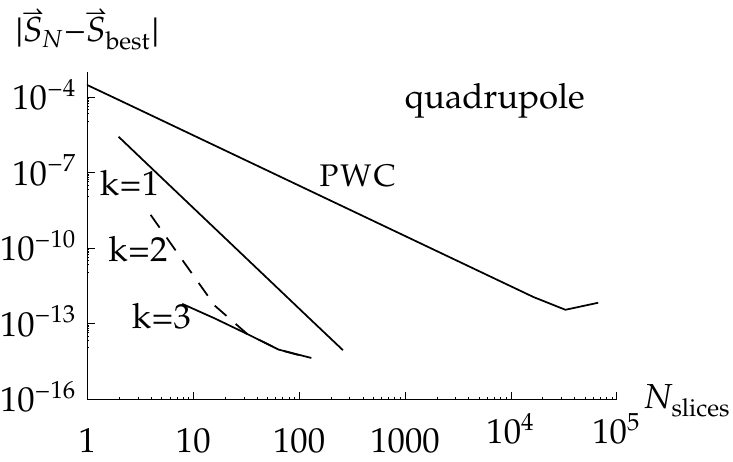}\quad%
  \includegraphics[scale=0.825]{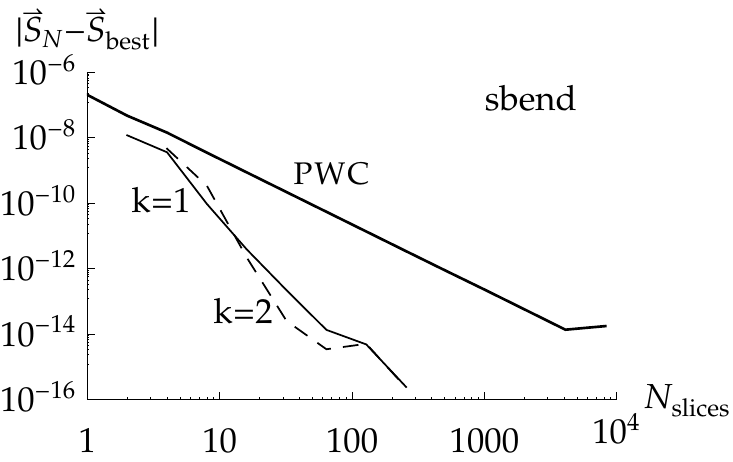}\\[2ex]
  \includegraphics[scale=0.825]{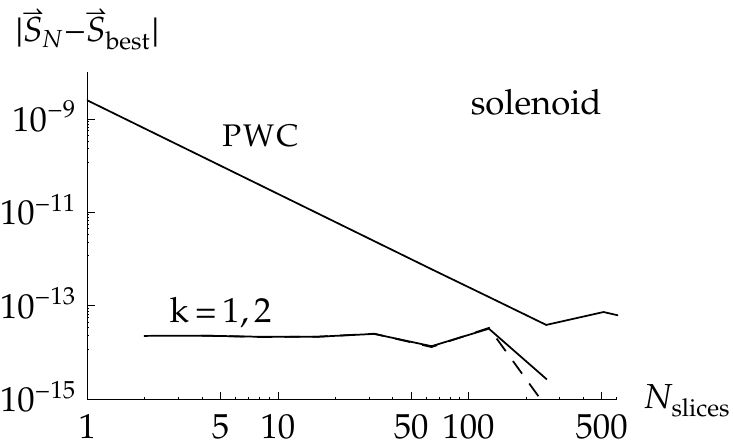}\quad%
  \includegraphics[scale=0.825]{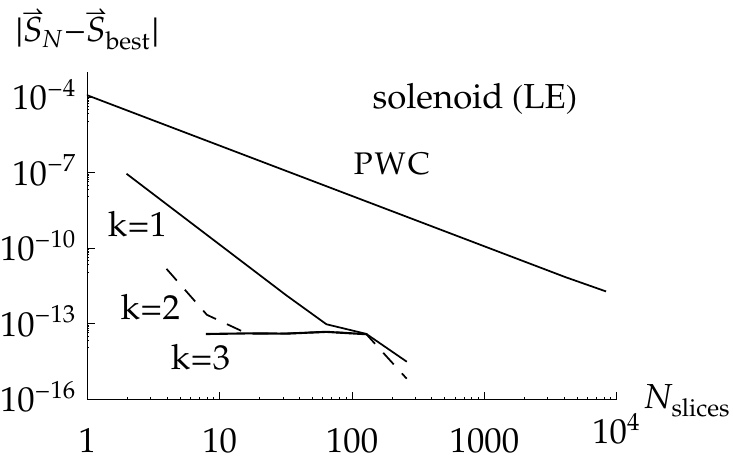}
  \caption{Absolute error in the computed spin versus the number of
    slices for a proton crossing several different beam\-line
    elements. All but the lower right graphic show results for a
    \SI{200}{GeV} $p^+$ crossing a \rhic\ \ir\ quad (top left),
    a \rhic\ arc dipole (top right), and a \SI{2.1}{\meter},
    \SI{1.3}{\tesla} solenoid (bottom left). The bottom-right
    graphic shows the result for a \SI{25}{GeV} $p^+$ traversing the
    same solenoid. Each curve corresponds to a different method
    of integrating the spin: \pwc\ for piece-wise constant integration;
    the $k$ values indicate the number of Romberg iterations.}
  \label{fig:elementErrsVsSlices}
\end{figure}

\begin{figure*}
  \includegraphics[width=0.98\textwidth]{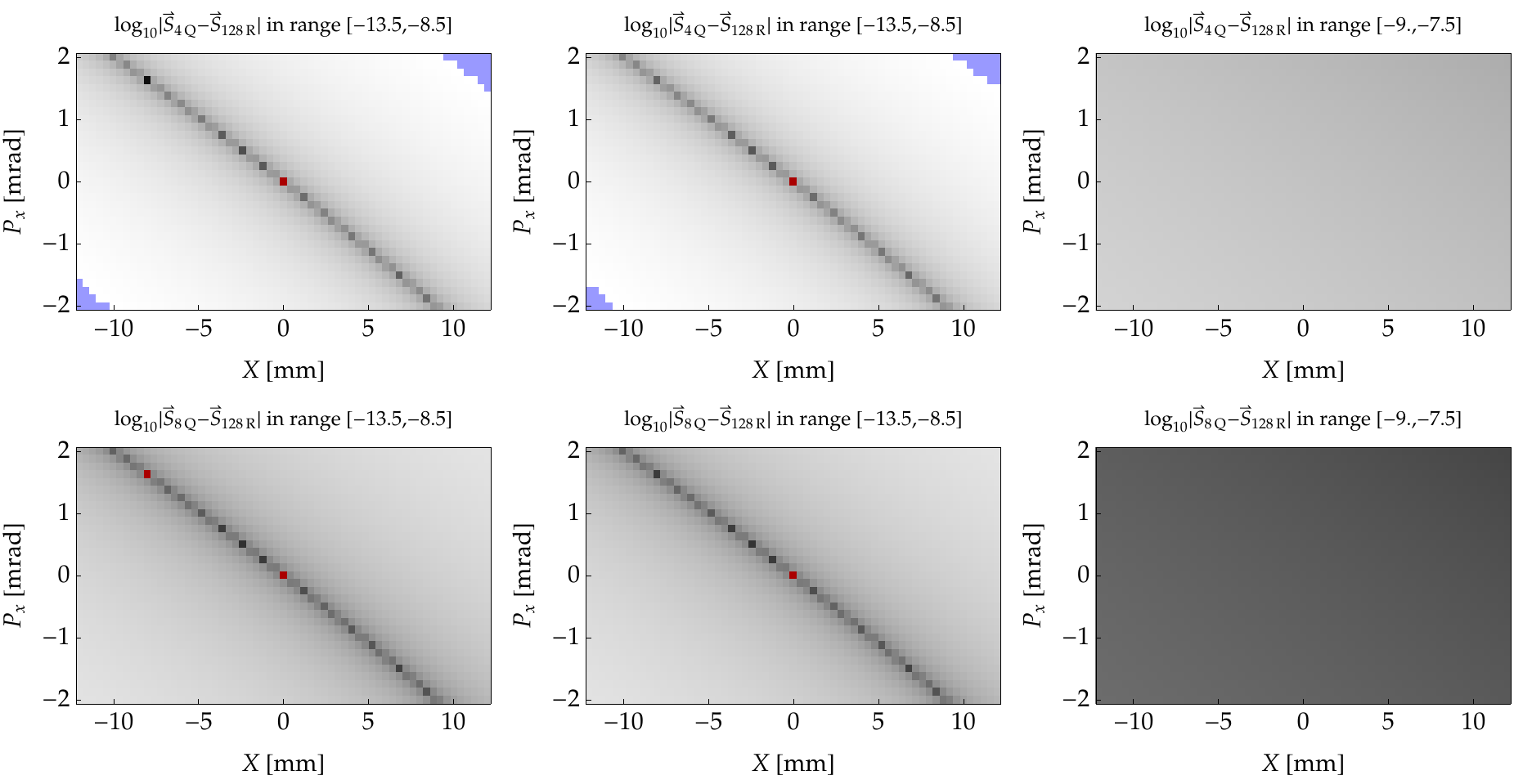}
  \caption{Absolute error in the computed spin across a \rhic\ dipole
for a range of initial conditions in orbital phase space. Darker
(lighter) grays represent smaller (larger) errors. Points outside
the range indicated above each graphic are colored red if below,
and blue if above. The error range covers the same five decades in
the four graphics to the left. In the right-hand two graphics, the
error range is again the same, but covers just a decade and a half
that is mostly above the range in the other graphics. The upper and
lower rows respectively show results obtained using \pwc\ spin
integration with four and eight slices. The three columns
correspond to different values for $\py^i$ and $\pt^i$: $\py^i=0$,
$\pt^i=0$ (left); $\py^i=\SI{20}{\murad}$, $\pt^i=0$ (middle);
$\py^i=\SI{20}{\murad}$, $\pt^i\approx\num{0.35e-3}$ (right).}
  \label{fig:sbendErrsQ}
\end{figure*}

The graphics in \refF{elementErrsVsSlices} show the spin integration
errors we computed when tracking a particular particle%
  \footnote{While not essential to our discussion, we note that
    this particle had the initial phase-space values $X^i = \SI{1.}{\mm}$,
    $\px^i = \SI{0.1}{\mrad}$, $\py^i = \SI{-0.2}{\mrad}$,
    and $\pt^i$ corresponding to a relative momentum deviation
    $\delta = \num{e-3}$. We set both $Y^i$ and $T^i$ to zero.
    In addition, we set this particle's initial spin to
    $\spin\approx(0.1952, 0.9759, 0.0976)$}
across several
different beamline elements. From top to bottom, those elements are
a quadrupole, a sector bend, a solenoid, and another solenoid (but
with a very low energy proton, such that the total spin rotation is
comparable to \ang{90}). In each of those graphics, the solid curve
labelled \pwc\ corresponds to piece-wise constant spin integration,
where for the orbital motion we used the integrators given in
\refSs{sbend}, \ref{sec:quad}, and \ref{sec:solenoid}. In all cases
the slope \num{-2} reveals second-order convergence for \pwc\ spin
integration. The remaining curves correspond to $k$ Romberg iterations
(\ie\ $Q_{kk}$) applied to the \pwc\ data obtained for the given
number of slices. Since we do not know the result of exact spin
integration, we have here estimated the spin integration error as
the absolute difference between the result $\spin^N$ obtained using
$N$ slices and what we considered our ``best'' result,
$\spin_\text{best}$.

\enlargethispage{\baselineskip}
In \refF{elementErrsVsSlices}, we have used for $\spin_\text{best}$
the results we obtained using \num{256} slices and three Romberg
iterations. If we instead use for $\spin_\text{best}$
our most finely sliced \pwc\ result, then small details in these
graphics change, but the overall implication holds---that one or more
iterations of the Romberg procedure can dramatically reduce the
errors made by \pwc\ spin integration. In the case of the quadrupole
integrated using four slices, we see that two Romberg iterations yield a
four-decade reduction in the error---to a level that requires some
four-hundred slices using just \pwc\ spin integration. For the solenoid,
a similar statement holds in the case of a low-energy proton; and
in the high-energy case, one Romberg iteration suffices to reduce the
spin integration error to the level of round-off.
For the sector bend, we see a less dramatic absolute reduction in
the error; but even there, between \num{8} and \num{64}~slices, the
slope \num{-4} on the $k=1$ curve tells us that one Romberg iteration
converts second-order \pwc\ integration to fourth-order.

\begin{figure*}
  \includegraphics[width=0.98\textwidth]{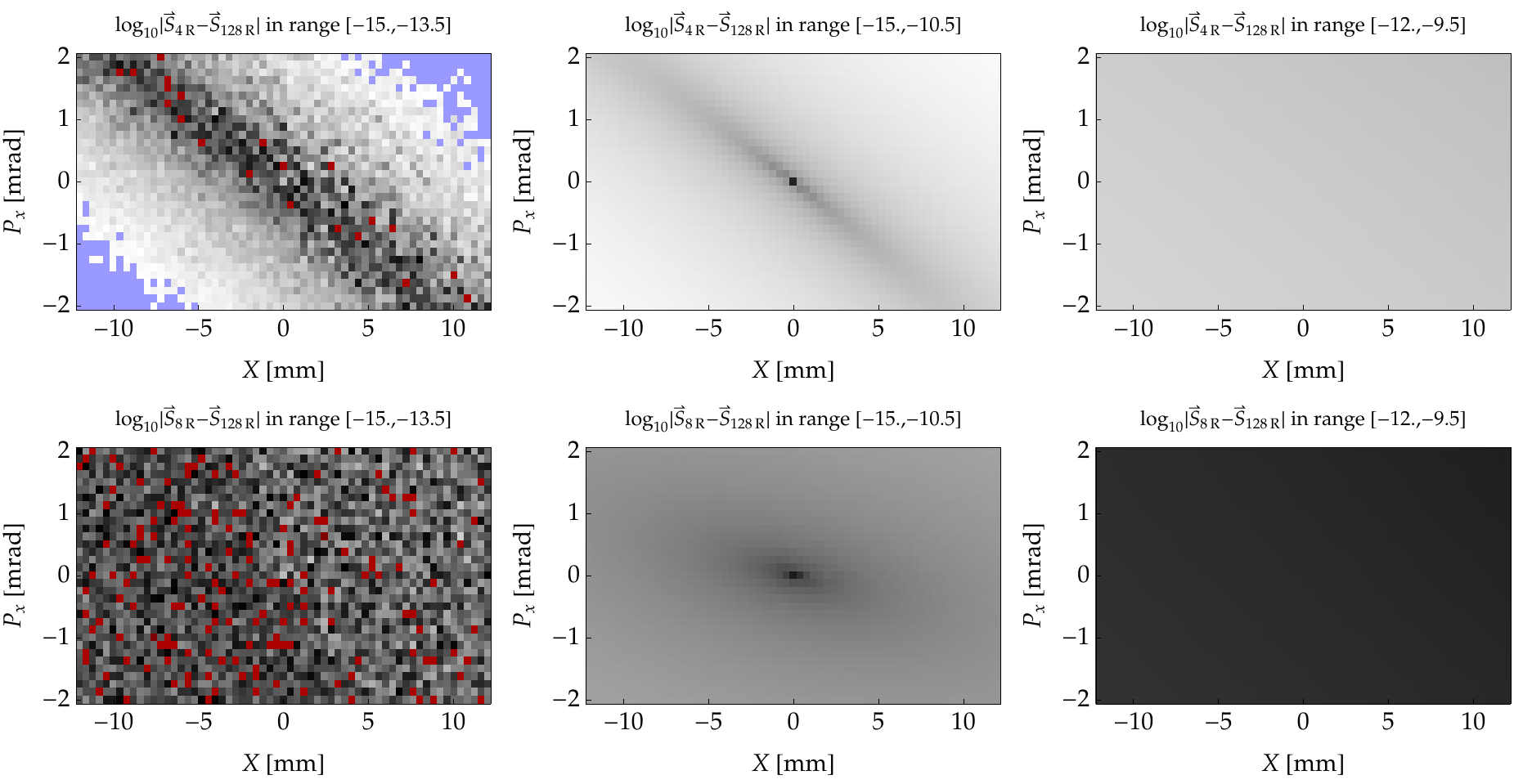}
  \caption{Absolute error in the computed spin across a \rhic\ dipole
for a range of initial conditions in orbital phase space. Each of the
graphics in this figure show results for the same orbital initial
conditions and the same number of slices as the corresponding graphic
in \refF{sbendErrsQ}. The differences are that the integration here
includes one Romberg step, and the error ranges are significantly lower.}
  \label{fig:sbendErrsR}
\end{figure*}

\begin{figure*}
  \includegraphics[width=\textwidth]{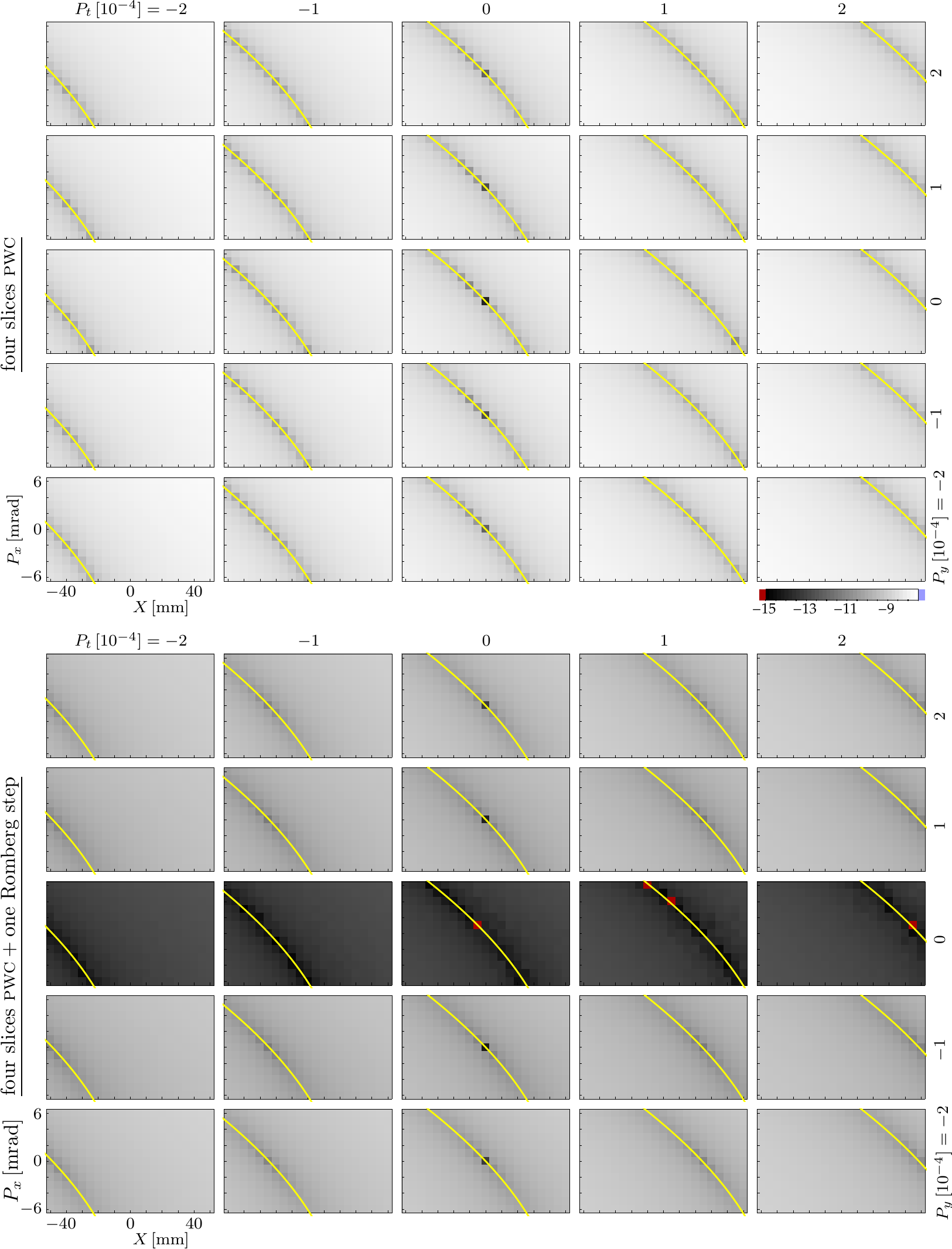}
  \caption{Absolute error in the computed spin using four slices
across a \rhic\ dipole. Each small
graphic covers a range of initial conditions in horizontal phase-space
(see labels at lower left). The rows and columns correspond
respectively to different initial values for $\py$ and $\pt$ (see
labels along right and top edges). For the lower set of graphics,
we added one Romberg iteration.}
  \label{fig:sbendErrsQR4M}
\end{figure*}

\begin{figure*}
  \includegraphics[width=\textwidth]{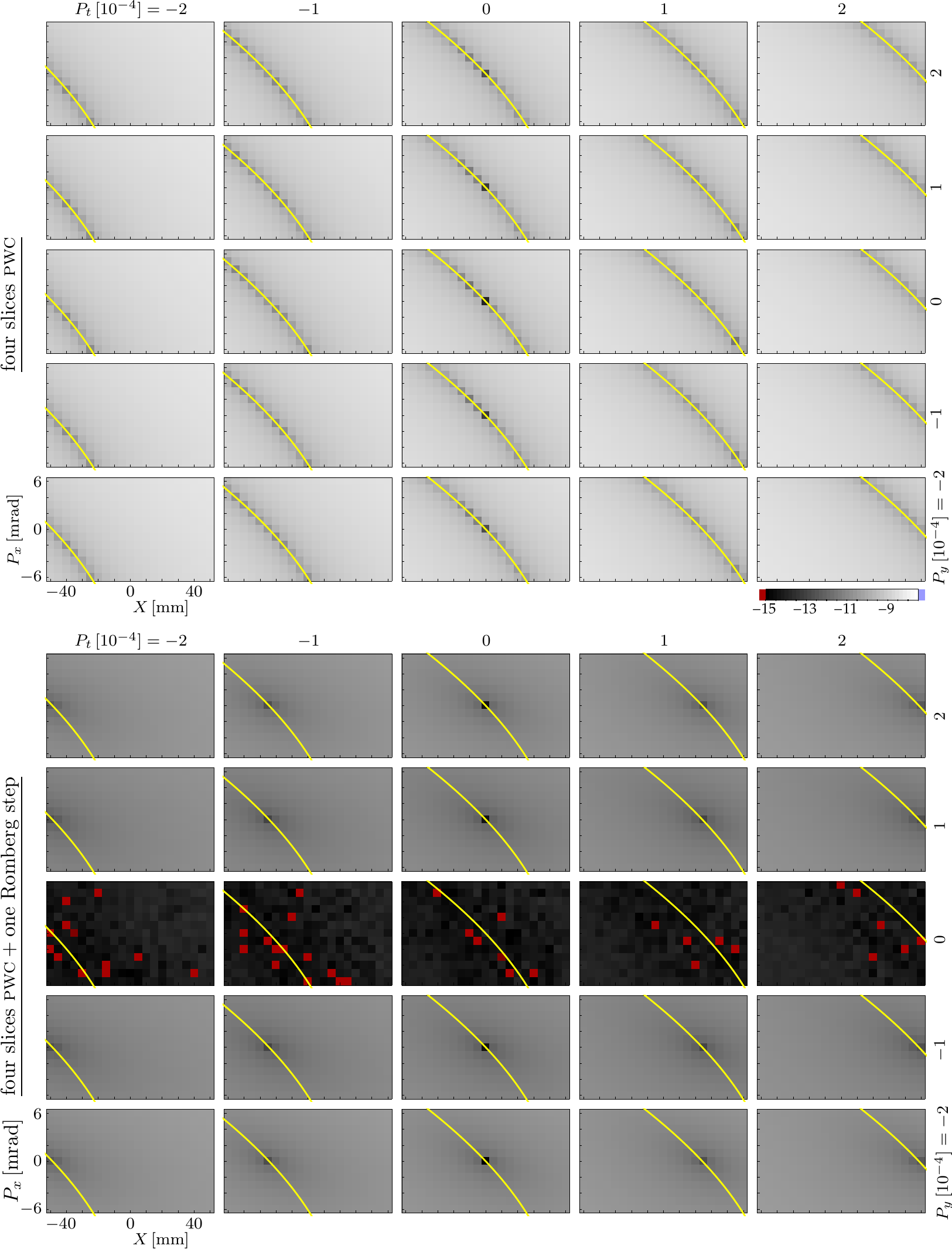}
  \caption{Absolute error in the computed spin using eight slices
across a \rhic\ dipole for a range of initial conditions. See also
the caption of \refF{sbendErrsQR4M}. The two sets of twenty-five
graphics correspond to \pwc\ spin integration (upper set) and
\pwc\ plus one Romberg iteration (lower set).}
  \label{fig:sbendErrsQR8M}
\end{figure*}

The graphics in \refF{elementErrsVsSlices} illustrate the scaling
of spin integration errors for one particular particle, and the
impact of Romberg iterations on \pwc\ spin integration. Were we
just lucky, or do we see similar behavior across the relevant
phase-space? The next several figures address this question,
illustrating the effect of Romberg quadratures on spin integration
for particles covering a range of initial conditions in phase-space.

For the sector bend, we examine spin integration errors across
a \rhic\ dipole: a \SI{9.45}{\meter} long magnet set to bend
\SI{200}{\GeV} protons \ang{2.225}, so with magnetic field strength
$B\approx\SI{2.75}{\tesla}$. As a proxy for exact tracking, we
settled on, after considerable testing, one Romberg iteration
applied to \num{128}~slices of \pwc\ spin integration.
Furthermore, because a magneto-static dipole is translationally
invariant in both $Y$ and $T$, we examine how the spin integration
errors vary with respect to just $X$, $\px$, $\py$, and $\pt$. 

In \refF{sbendErrsQ}, we show the absolute error in the computed
spin using \pwc\ spin integration with either four or eight slices
(upper and lower rows, respectively).
\refFc{sbendErrsR} shows the corresponding results obtained after
we apply one Romberg iteration. To help the reader make a proper
comparison, we note further details in the captions of those two
figures.

A comparison of \refFs{sbendErrsQ} and \ref{fig:sbendErrsR}
suggests that applying Romberg quadratures to spin integration
across a sector bend does indeed reduce the error throughout
phase-space. But we are also struck by some of the features
seen in \refF{sbendErrsQ}: the pronounced valley that runs
diagonally across the four graphics on the left; and the
much flatter and somewhat larger errors in the presence of an
energy deviation, seen in the two graphics on the right.

To gain some understanding of these features, we examined a much
larger range of phase space; a range that is, in fact un-physically
large for an arc dipole in \rhic. \refFsc{sbendErrsQR4M} and
\ref{fig:sbendErrsQR8M} show the results. Each of the one hundred
small graphics in those two figures covers the same range of
horizontal phase space: \SI{\pm5}{\cm} in $X$, and
\SI{\pm6}{\mrad} in $\px$. They are laid out in four matrices of
twenty-five graphics each, where the columns correspond to five
different values of $\pt^i$, evenly spaced in the range \num{\pm2e-4};
and the rows correspond to five different values of $\py^i$, evenly
spaced in the range \SI{\pm0.2}{\mrad}. (See the labels at the top
and to the right of each matrix of graphics.) To obtain the graphics
in \refF{sbendErrsQR4M}, we used four slices of \pwc\ spin integration
for the upper set of twenty-five graphics, and four slices plus
one Romberg iteration for the lower set. We obtained the data shown
in \refF{sbendErrsQR8M} using eight slices. In all the graphics,
the gray-scale covers the same (logarithmic) range of errors:
from black at \num{e-15} to white at \num{e-7.5}.

The graphics shown in \refFs{sbendErrsQR4M} and \ref{fig:sbendErrsQR8M}
reiterate the message that applying a Romberg step to the result
of \pwc\ spin integration can reduce the error throughout the
phase-space of orbital initial conditions. In addition, however,
we draw the reader's attention to the yellow curve, overlaid on
each graphic, which clearly follows the deep valley seen in those
graphics for which no Romberg step was applied. In each graphic,
this curve is the locus of points $(X^i,\px^i)$ for which the
value of $\px$ at the magnet exit equals the value $\px^i$ at
the magnet entrance. This value is given by
\begin{multline}\label{eq:dpx0}
  \px^\text{sym}(X^i) = \cos\Bigl(\frac{hL}{2}\Bigr)
    \biggl[\sqrt{1 + \frac{2}{\smash{\bo}}\pt + \pt^2 - \py^2
                   - (1 + hX^i)^2\cos^2\Bigl(\frac{hL}{2}\Bigr)}\\
            - (1 + hX^i)\sin\Bigl(\frac{hL}{2}\Bigr)\bigg].
\end{multline}
Note that the dependence of $\px^\text{sym}$ on $\py$ and $\pt$ occurs
only under the square-root, and that there it is linear in $\pt$ and
quadratic in $\py$. This difference (together with the fact that
$\py$ and $\pt$ denote small quantities) explains why the yellow curves
in \refFs{sbendErrsQR4M} and \ref{fig:sbendErrsQR8M} show little
dependence on $\py$, but very sensitive dependence on $\pt$.

Earlier, when examining \refF{sbendErrsQ}, we noted that the
two right-hand graphics have much flatter and somewhat larger
error profiles than appear in the left-hand four graphics.
Now, on comparing with the graphics in \refF{sbendErrsQR4M},
we see what happened: the valley simply moved ``off stage''
as we increased the energy deviation.

\begin{figure*}
  \includegraphics[width=0.97\textwidth]{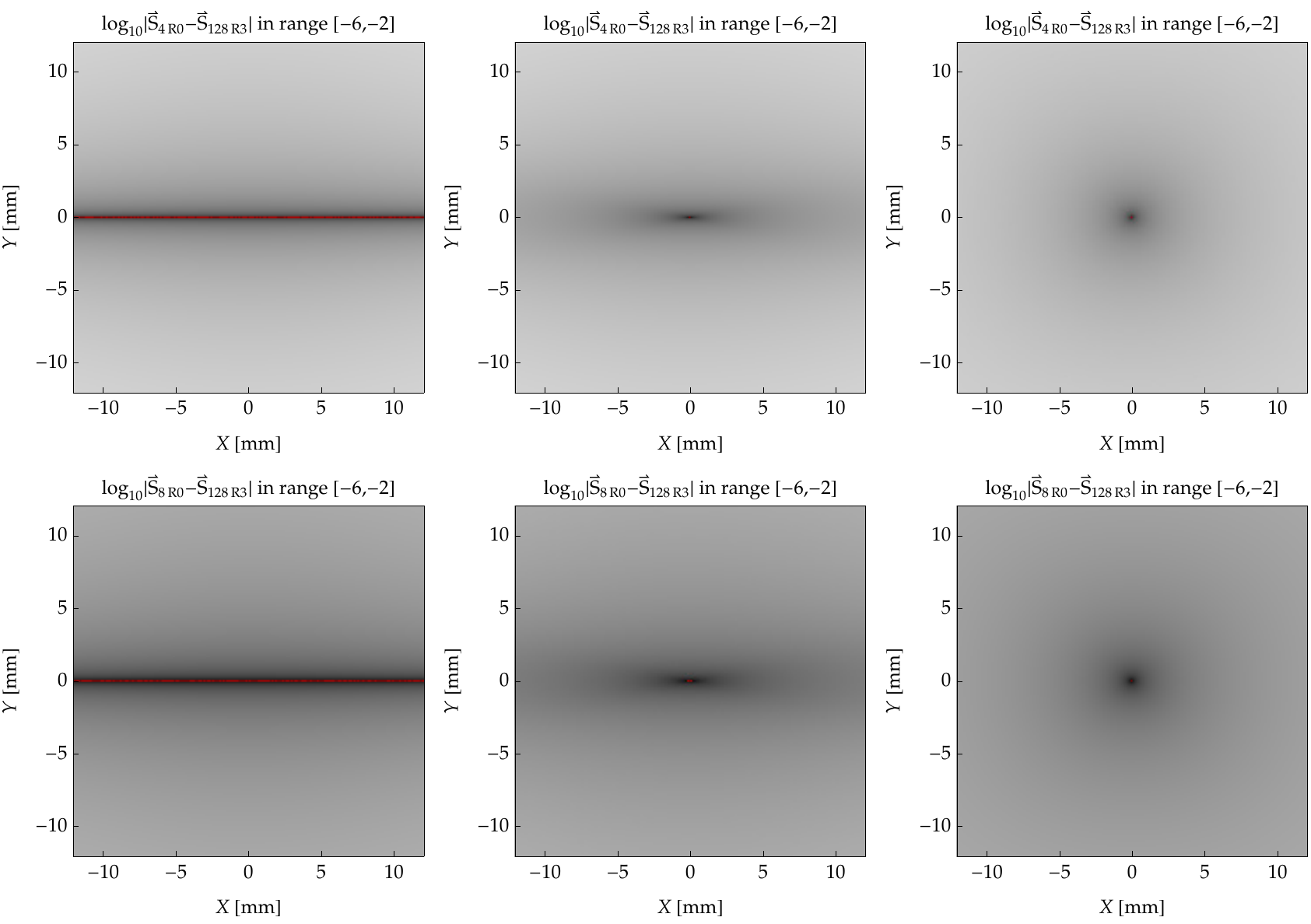}\\[2ex]
  \includegraphics[width=0.97\textwidth]{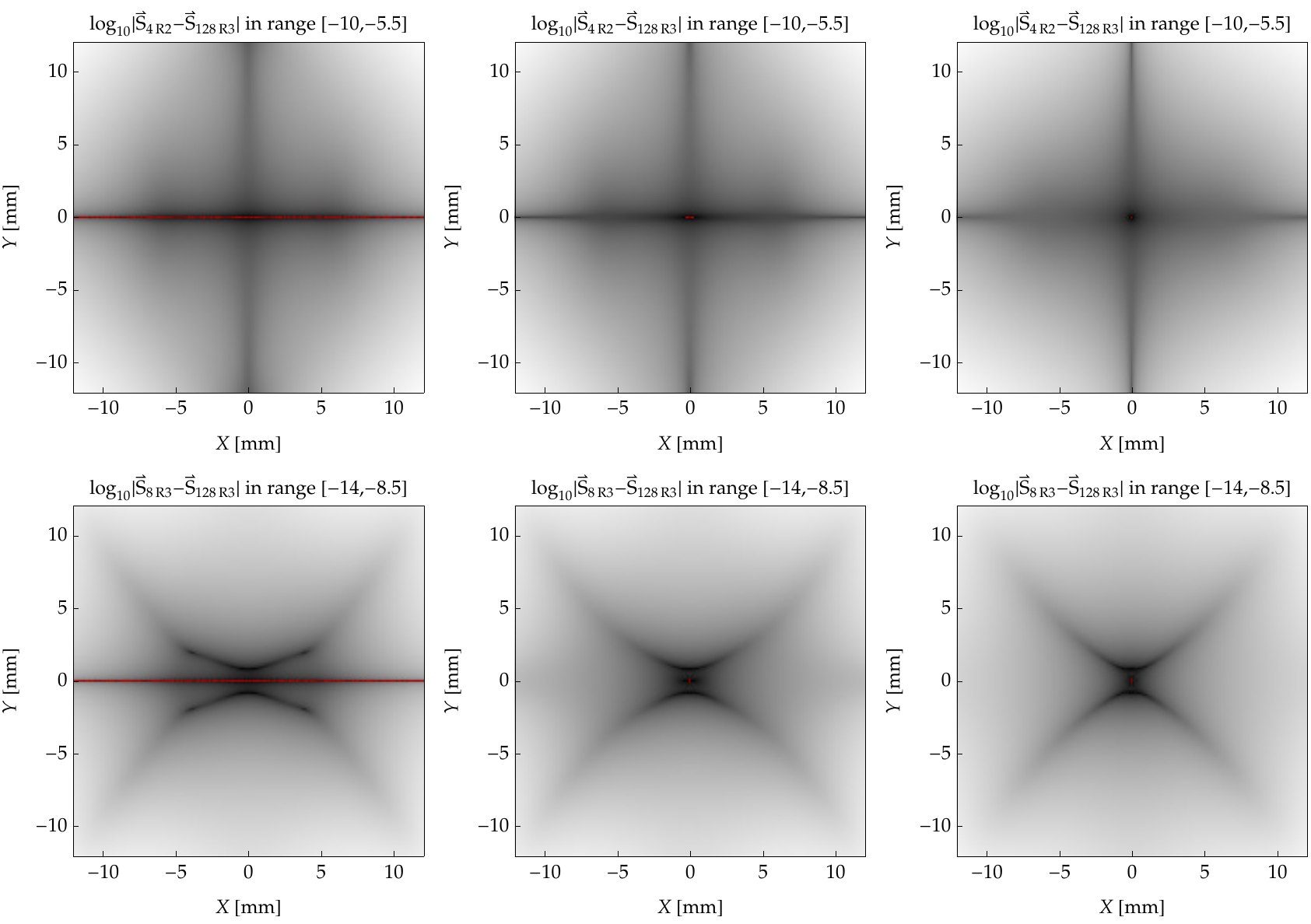}
  \caption{Absolute error in the computed spin across a
\rhic\ \ir\ quadrupole. The rows, from top to bottom, show results
obtained using four slices, eight slices, four slices plus
two Romberg steps, and eight slices plus three Romberg steps.
The columns, from left to right, show results averaged over spins
covering an opening angle about the vertical of \ang{0}, \ang{20},
and \ang{180}.}
  \label{fig:quadErrs48QR}
\end{figure*}

For the quadrupole, we examine spin integration errors across
a \rhic\ interaction region (\ir) quadrupole: in this case a
\SI{1.83}{\meter} long magnet set to focus \SI{200}{\GeV} protons
with a magnetic quadrupole gradient of $b_2=\SI{70.0}{\tesla\per\meter}$.
As a proxy for exact tracking, we settled on, again after much
testing, three Romberg iterations applied to \num{128}~slices
of \pwc\ spin integration.

A coarse examination over the range of initial conditions
$\SI{\pm20}{\mm}$ in $X$ and $Y$,
$\SI{\pm20}{\murad}$ in $\px$ and $\py$, and
$\num{\pm0.35e-3}$ in $\pt$
indicate that (at least within this phase-space domain) spin
integration errors change little with respect to variations in
the transverse momenta $\px$ and $\py$, or the relative energy
deviation $\pt$. As a consequence, here we show graphics of the
spin integration errors for this element only in the $(X,Y)$
plane, with $\px^i=\py^i=\pt^i=0$.

In the case of the sector bend, the spin integration error does
not appear to depend on the initial spin orientation. For the
quadrupole, however, the accuracy of spin integration very
definitely depends on the initial spin, so we have included
that dependence in the results shown here.

In \refF{quadErrs48QR}, we show the absolute error in the
computed spin across a \rhic\ \ir\ quadrupole. The first two rows
show results obtained with \pwc\ spin integration using either
four slices (first row), or eight slices (second). The results
shown in the lower two rows correspond to applying our Romberg
procedure to the first two rows: four slices plus two Romberg
steps (third row), or eight slices plus three Romberg steps (bottom row).
As indicated above each graphic, the gray-scale in the first two
rows covers the same (logarithmic) range of errors: from black at
\num{e-6} to white at \num{e-2}. In the third row, the range goes
from \num{e-10} to \num{e-5.5}, which means most of the errors in
the third row (all except small triangles in the corners) lie below
the errors in the upper two rows. In the last row, the range of
\num{e-14} to \num{e-8.5} means the errors here lie entirely below
those of the upper two rows; and they lie below most of the errors
in the third row.

The columns in \refF{quadErrs48QR} correspond to averaging over
various sets of initial spins. We generated a quasi-random
distribution of spins (\ie, points on the unit sphere%
  ~\cite{Marsaglia:1972:ChoosePtSphere})
with opening angle $\alpha$ about the vertical:
$\alpha=\ang{0}$ (left column), \ang{20} (middle), and \ang{180}
(right).

In the case of a vertical spin ($\alpha=\ang{0}$), the errors
vanish for orbital motion in the mid-plane because the vertical
spin remains parallel to the vertical field everywhere along
those orbits and therefore does not precess. What about other
particles, those launched still with a vertical initial spin,
but away from the mid-plane? For particles traversing a
\rhic\ \ir\ quad, the $\bigOmega$ of \refE{t-bmt.so} lies
predominantly along the quadrupole magnetic field, $(b_2Y,b_2X,0)$.
Since the component of $\bigOmega$ orthogonal to the spin
is the component that rotates the spin, it follows that an
initially vertical spin should be most sensitive to variations
in $Y$, and rather insensitive to variations in $X$. This
suggests why the two graphics in the upper left of
\refF{quadErrs48QR}---those for $\alpha=\ang{0}$ and no
Romberg iterations---show essentially no variation with respect
to $X^i$. Moreover, the fact that the orbital variation in $Y$
is (to first order) proportional to $Y^i$ suggests why the
errors in those same two graphics increase away from the
horizontal mid-plane.

As we allow the distribution of initial spins to open up
($\alpha>\ang{0}$), the distribution of errors (before the
application of any Romberg iterations) quickly becomes more
symmetric about the quadrupole axis. See the four graphics
in the upper-right of \refF{quadErrs48QR}. After we apply one
or more Romberg steps, the quadrupole spin-integration errors
become smaller and develop additional structure across the
$(X,Y)$ plane. See the bottom two rows of \refF{quadErrs48QR}.
We do not yet understand the origin of this structure.

\begin{figure}
  \includegraphics*[width=0.32\textwidth,trim= 0 340 0   0]{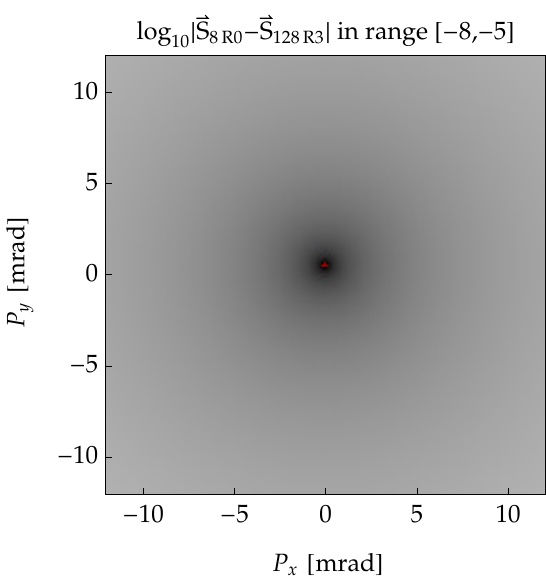}
  \ %
  \includegraphics*[width=0.32\textwidth,trim= 0 170 0 170]{fig13_solenoid}
  \ %
  \includegraphics*[width=0.32\textwidth,trim= 0   0 0 340]{fig13_solenoid}
  \caption{Absolute error in the computed spin across a solenoid.
We computed the data shown in these three graphics using, from left
to right, eight slices \pwc\ spin integration, four slices \pwc\ plus
two Romberg iterations, and eight slices \pwc\ plus three Romberg
iterations. Note the different error ranges indicated above each
graphic: with each step across the page, the errors fall by several
decades.}
  \label{fig:solenoidErrs48QR}
\end{figure}

For the solenoid, we examine spin integration errors across a
\SI{1.3}{\tesla}, \SI{2.1}{\meter} long magnet. For the \SI{200}{\GeV}
protons we  used in previous elements, we find that four slices together
with one or two levels of Romberg quadratures suffice to reduce errors
to the level of round-off. In this case, however, the particle
experiences very little spin rotation---of order \SI{10}{\mrad}.
We therefore chose to look at the spin integration errors for a
\emph{much} lower energy proton, \SI{1.2}{\GeV}, traversing this
same magnet. In this case the spin rotation is of order
\SI{0.6}{\radian}. As a proxy for exact tracking, we again
settled on three Romberg iterations applied to \num{128}~slices
of \pwc\ spin integration.

Given the homogeneous nature of a solenoid's body magnetic field
in our thin-fringe approximation, we expect spin integration errors
across this element to exhibit very little dependence on
$X^i$, $Y^i$, or $\pt^i$. A coarse examination over a range
of initial conditions in phase space indicates that this is
indeed the case. As a consequence, here we show graphics of
the spin integration errors for this element only in the
$(\px,\py)$ plane.

In \refF{solenoidErrs48QR}, we show the absolute error in the
computed spin for a \SI{1.2}{\GeV} proton traversing the solenoid
described above. The top graphic shows the result obtained
using eight slices of \pwc\ spin integration. The two graphics
below show results obtained using four slices plus two Romberg
steps (middle), and eight slices plus three Romberg steps (bottom).
As indicated by the ranges noted above each graphic, the error
ranges in these three graphics do not overlap. Indeed, the error
range drops by several decades with each step down the page.

\enlargethispage{-\baselineskip}
\subsubsection{Single-turn errors}

\begin{figure}
  \includegraphics{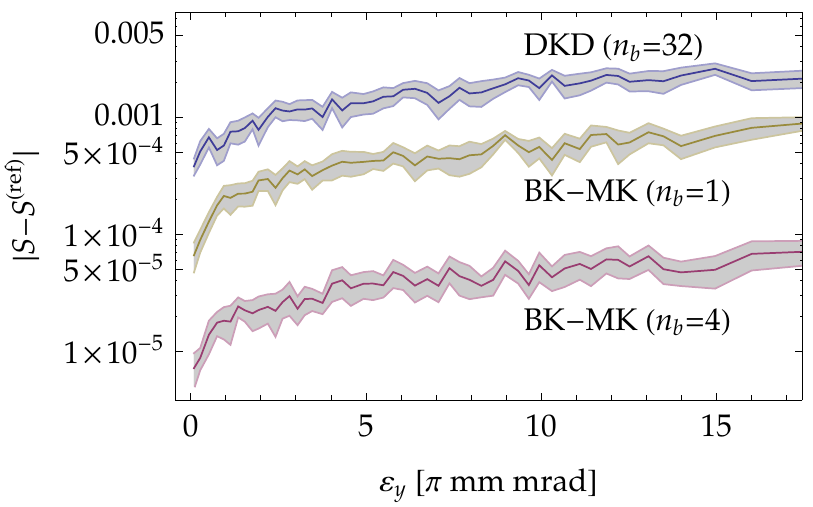}
  \caption{Amplitude dependence of the single-turn spin errors for
three different integrators: drift-kick with $n_b = 32$ (upper),
and \bkmk\ with $n_b = 1$ (middle), and $n_b = 4$ (lower).}
  \label{fig:spinErrsVsAmp}
\end{figure}

In this section, we examine spin integration errors after a full
turn in the same \rhic\ lattice used for the orbital studies in
\refS{orbitPerform}.
See \refFs{oneTurnSpinErrs} and \ref{fig:spinErrsVsAmp}.
As our measure of the error, we use the mean difference between
the spin computed using a given set of numerical algorithms, and a
reference solution that we use as a proxy for the exact result.
The mean is taken over an ensemble of spins in a beam. We compute
thus, at turn $t$,
\begin{equation}
  \Delta\spin(t) = \frac{1}{N}\sum_{j=1}^N\,
    \Bigl|\spin_j(t) - \spin_j^\text{\,(ref)}(t)\Bigr|\,.
\end{equation}
For the results shown here, we computed the reference solution
using \bkmk\ integration with $n_b = 256$ and three Romberg
iterations.

\refFc{oneTurnSpinErrs} shows the absolute spin error as a function
of the number of spin slices, here chosen identical to the number
of orbital slices. No Romberg steps were applied, except when
computing the reference solution. As a consequence, differences
seen here between drift-kick results (upper two curves) and
\bkmk\ results (lower curve) derive solely from differences in
the orbital data passed to the \tbmt\ equation.

\begin{figure}
  \includegraphics{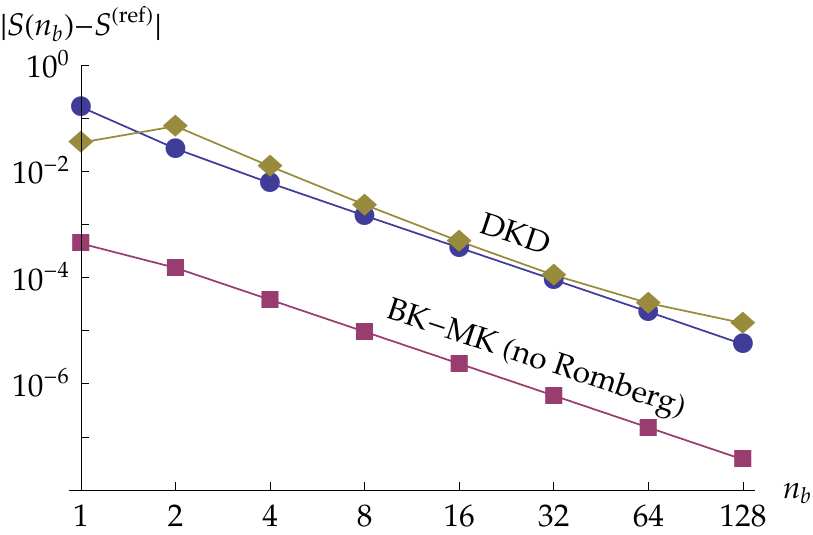}
  \caption{Single-turn spin errors as a function of the number of
slices for several different integrators: drift-kick (blue circles,
tan diamonds), and \bkmk\ (purple squares). The blue circles
represent \dks: traverse each slice using half-drift,
half-kick, spin rotation, half-kick, half-drift. The purple squares
represent \dkds: traverse each slice using half-\dkd,
spin rotation, half-\dkd. The reference solution used \bkmk\ 
with $n_b = 256$ together with three Romberg iterations.}
  \label{fig:oneTurnSpinErrs}
\end{figure}

At small numbers of orbital slices, spin integration based on drift-kick
orbital integration reproduces only the one-turn spin-precession
axis (close to the vertical). The precession angle about that axis,
however, is completely off: compared to the reference solution, the
phase of the spin precession is distributed uniformly over a $2\pi$
interval. Increasing the number of orbital slices improves the
accuracy of spin integration, and at $n_b = 64$ the single-turn
spin-phase errors fall below a \si{\mrad}.

When performing spin integration based on \bkmk\ orbital
integration, we obtain errors that are consistently two-and-a-half
decades below those based on drift-kick integration (see the lowest
curve in \refF{oneTurnSpinErrs}). This result, we emphasize, obtains
in the \emph{absence} of Romberg quadrature.

\refFc{spinErrsVsAmp} also shows the absolute spin error, but now
as a function of particle emittance, after a single turn in the
\rhic\ lattice. As in \refF{oneTurnSpinErrs}, no Romberg steps
were applied---except in computing the reference solution---so
that here also the differences seen derive solely from differences
in the accuracy of the orbital integration. In addition, we have
reduced the noise level in this graphic by, as in \refF{xErrVsAmp},
averaging over narrow slices of emittance.

\subsubsection{Evolution of spin errors}

\begin{figure}
  \includegraphics{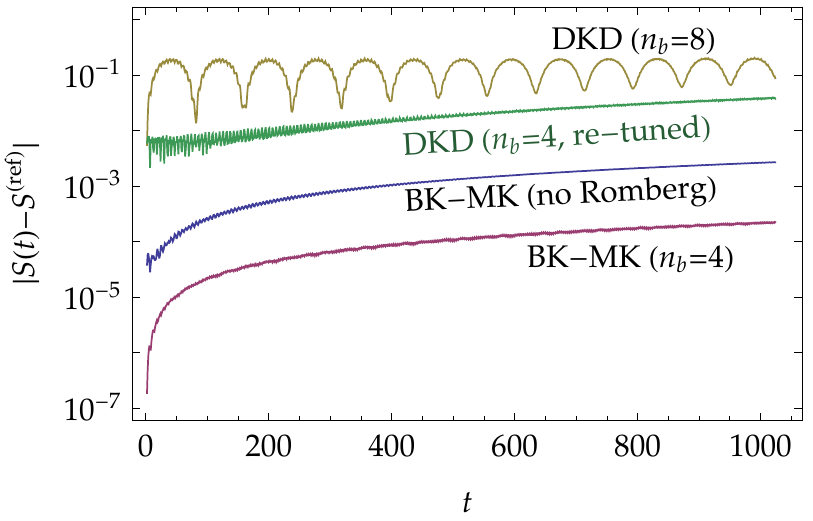}
  \caption{Accumulated spin error as a function of turn number, for
multi-turn tracking in a \rhic\ lattice, using different methods of
integration. From top to bottom:
drift-kick integration with $n_b = 8$ and no tune-fit (tan);
the previous standard, a \teapot\ split with $n_b = 4$ (green);
\bkmk\ with $n_b = 4$ and no Romberg (blue);
and \bkmk\ with $n_b = 4$ (purple).
The simulated beam had a \SI{95}{\percent} emittance of
\SI{15\pi}{\mm\,\mrad}.}
  \label{fig:spinErrVsTurns}
\end{figure}

We also examined how spin errors evolve over the course of many
simulated turns in \rhic. In \refF{spinErrVsTurns} we use the
accumulated spin error to compare different methods of integration.
The green curve shows results obtained using the previous standard:
a \teapot\ split with four slices for most elements, but $16$~slices
for the strong elements around the interaction regions. The lowest
two curves we obtained using the new integrators, also with $n_b=4$.
For the blue curve, we used just the ``trapezoidal'' rule
(\pwc\ of \refF{elementErrsVsSlices}). For the lowest curve, we computed
a Romberg limit using a maximum $k$ of \num{3}.

The topmost curve, obtained using drift-kick integration with
$n_b=8$, but without retuning the lattice, exhibits pronounced
oscillations. The tune errors in the orbital integration translate
into these periodic oscillations of the spin errors.

\section{\gpu-accelerated implementation}
\label{sec:GPUimplementation}

In recent years, graphics processing units (\gpu{}s) have emerged
as an effective means to accelerate compute-intensive workloads.
Because they are designed for massively data-parallel workloads with
very regular control flow and memory access patterns, \gpu{}s can
dedicate a much larger fraction of their transistors to processing
elements (registers and arithmetic and logic units) than can \cpu{}s.
As a consequence, they can achieve significantly larger computational
throughput, performance per watt, and performance per dollar for
workloads that are well-suited to their architecture. Particle
tracking in the absence of space-charge effects is an embarrassingly
parallel problem, and is therefore a natural fit for \gpu{}s.

To map the spin-orbit tracking problem onto \gpu\ hardware, we
assign a \gpu\ thread to each particle. For the actual tracking,
we first transfer the state of a bunch to global memory on the \gpu.
This state comprises six-dimensional orbit data, three-dimensional
spin data, plus the quaternions required for accumulating spin rotations.
Particle tracking through individual beamline elements is broken up into
several \gpu\ kernels. We first compute the orbital motion through
a particular beamline element. This requires loading orbit data at
the element entrance from global memory into registers, computing
the orbital motion using the integrators discussed in detail above,
and then, at the element exit, writing the updated state back to
global memory. During the orbital computation, we record in global
memory the spin precession vector $\bigOmega$ at the edge of each orbital
slice. After integrating the orbital data through the given element,
we then use the recorded $\bigOmega$ data to update the quaternions
using \pwc\ spin integration plus one or more Romberg steps. We then
update the spin.

The particle data resides on the \gpu\ throughout spin-orbit tracking.
By maintaining the data on the \gpu, we avoid unnecessary data transfers
between \gpu\ and \cpu, which can significantly slow down simulations.
Particle data is occasionally copied back to the \cpu\ in order to
perform analysis; \eg, to compute (or update) the invariant spin field
(\isf) at a particular location in the accelerator.

\begin{figure}
  \includegraphics[scale=1.05]{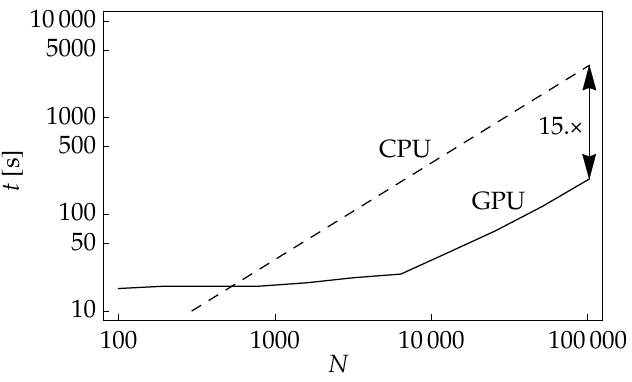}
  \caption{Time required to track $N$~particles through 100~turns
of the \rhic\ lattice with \gpu-accelerated integrators (solid). The
\cpu\ time (dashed) is an estimate; see text for details.}
  \label{fig:singleGPUScaling}
\end{figure}

\refFc{singleGPUScaling} shows a comparison of the performance of
spin-orbit tracking on a \gpu\ and a \cpu. The \gpu\ simulations
were carried out on the \textsc{Dirac} cluster at the National
Energy Research Scientific Computing center (\textsc{NERSC}).
The \cpu\ performance is an estimate for an \num{8}-core \cpu,
based on the assumption that the simulation scales perfectly to
\num{8}~\cpu\ cores and that each \cpu\ core is a factor \num{4}
faster than a \gpu\ core. For up to a few hundred particles the
\cpu\ is faster than the \gpu. For larger numbers of particles, the
\gpu\ accelerated simulations are faster. For a few ten-thousands
of particles, the \gpu\ accelerated simulations are approximately
\num{15} times faster than the corresponding \cpu\ implementation.  

\section{Conclusion}
\label{sec:conclusion}


In \refS{orbitPerform}, we showed that the \bkmk\ integrators can
be much more efficient than the drift-kick integrators. This, of
course, is well known (but see Forest's discussion%
    ~\cite{Forest:2006:GeomIntegPA}
for why drift-kick integration remains useful).
Our purpose in including such a comparison here has to do with
the influence of accurate orbit integration on the acccuracy of
spin integration.

\refFc{oneTurnSpinErrs} tells us that better orbit data yields more
accurate spin integration. Using \bkmk\ integration yields a two-decade
reduction of the spin errors as compared to drift-kick integration.
We see that the improvement with the number of orbital slices is,
in large part, due to the improvement in the orbital accuracy.
In other words, with drift-kick integration the accuracy of the
spin motion will improve just by increasing the number of orbital
slices while holding the number of spin integration steps fixed.

On the other hand, for \bkmk\ integration, the discretization of the
spin motion appears to be a significant source of error:
A comparison of \refFs{oneTurnErrsX} and \ref{fig:xErrVsAmp}
with \refFs{oneTurnSpinErrs} and \ref{fig:spinErrsVsAmp} show
that the relative error in the orbital motion is much smaller
than that in the spin rotation angle.

In \refS{bezier} we discussed interpolating the spin precession
vectors, using cubic B\'ezier polynomials, as a means of rapidly
computing multiple spin steps per orbital slice. To compute a
B\'ezier curve for $\bigOmega$, one must first determine the
associated control points (see \refF{cubicBezier}), and
doing so requires computing the longitudinal derivative of
$\bigOmega$ at the beginning and end of each slice. For most
elements, this is a non-trivial, hence time-consuming, computation.
In other words, if we wish to enjoy the fast computation of many
$\bigOmega$\hairsp{s} along an interpolating B\'ezier curve, we
must first invest some time computing $\bigOmega'$. As a trade-off,
this means we benefit only if we plan to interpolate $\bigOmega$
at some (possibly large) number of intermediate locations.

In the bargain, however, we do not improve the \emph{rate} at
which spin integration converges: The accuracy of this approach
cannot improve upon what we obtain by taking smaller steps
(\ie, using more orbital slices). In other words, the best we
might hope for from interpolating the precession vector
$\bigOmega$, is a faster computation of \pwc\ spin integration,
which exhibits just second-order convergence.


As shown in \refF{elementErrsVsSlices}, as well as in
\refFs{sbendErrsQ} through~\ref{fig:solenoidErrs48QR}, the
application of Romberg quadratures to \pwc\ spin integration
\emph{does} increase the rate of convergence. Moreover, this
approach requires only a modest number of orbital slices.
Applying the Romberg algorithm to a sequence of quaternion
products constitutes an easy and fast means of accelerating
the convergence of spin integration. As a consequence, we have
not pursued further the use of B\'ezier curves to speed up
the computation of $\bigOmega$.

Note that our \emph{initial} motivation for using quaternions
was the factor of two savings in arithmetic they convey. After
discovering that Romberg applied to quaternions accelerates the
convergence of the spin motion, we have another---much more
significant---reason to use quaternions when integrating the
\tbmt\ equation for spin motion.


Concerning the convergence of spin motion, we here comment
on what happens when simulating a system near a spin-orbit
resonance%
  ~\cite{Barber:2004:Qperiodic}.
First, note that in such a system a particle's spin orientation
becomes very sensitive to even very tiny perturbations.
As a consequence, if we follow a single spin and ask for
it's motion to converge as we slice the
lattice ever more finely, we may never see that happen.
As with the orbital motion, we do not---cannot---ask for
the ``exact result'', because the machine we build is not
the one we designed (though we hope it is close!). We
instead ask for accurate qualitative dynamics, as revealed
by phase-space portraits.

Second, note that the integrators are particular to given
elements. In other words, they care only about the mapping
of phase-space co\"ordinates across a single element:
  $z^i\text{ at entrance} \rightarrow z^f \text{ at exit}$.
On the other hand, a resonance condition can exist only in
a ring: it is \emph{defined} by the ring. One might therefore
expect that the choice of integrator is, in some sense,
independent of the proximity and strength of a resonance.
But this expectation can hold \emph{only if the integrators
are independent of, or have no effect on, whatever properties
define the resonance.}

When using drift-kick integrators, we know the tune seen in
tracking data depends on how finely we slice our elements.
What happens, then, as we approach a resonance? If we increase
the slicing so as to ensure convergence, the effective tune
changes. Hence the distance to resonance changes, and the
spin dynamics can appear quite different---especially because
of its sensitivity in the neighborhood of a resonance.
If, on increasing the slicing we also re-tune the lattice,
then the distance to resonance, as defined by the orbital
tunes, remains the same. But the quadrupole strengths do
change, and this implies a change in the spin tune for a
given simulated particle%
  ~\cite{Barber:2004:Qperiodic}.
And so again the distance to resonance changes, and again
the spin dynamics can appear quite different because of its
sensitivity in the neighborhood of a resonance.

Because \bkmk\ integrators yield the same linear orbital tunes
independently of slicing, they are not subject to the same
difficulties.

In addition to the improved accuracy and accelerated convergence
of spin tracking, our code benefits from the significant speed-up
it derives just from its implementation on \gpu{s}. As shown in
\refF{singleGPUScaling}, a reasonable estimate for that benefit
yields a speed-up factor of about 15. On a more practical level,
we note that on one of \textsc{NVIDIA's} \textsc{GeForce TITAN} \gpu\ nodes,
computing an \isf\ for \rhic\ at \num{e4}~phase-space locations
using \num{e3}~turns takes less than fifteen minutes.

Finally, we are currently implementing integrators for other types of
optical elements. These include electrostatic lenses, which are important
for simulating \textsc{EDM} experiments, as well as better models for
Siberian snakes.

\section*{Acknowledgments}

We thank Dr.\,Javier von Stecher for useful discussions and his help
with some of the figures.
This work is supported by the U.S. Department of Energy,
Office of Science, Office of Nuclear Physics, including grant
No.~DE-SC0004432.
In addition, this research used resources of the National Energy
Research Scientific Computing Center (\nersc), which is supported
by the Office of Science of the U.S. Department of
Energy under Contract No.~DE-AC02-05CH11231.

\providecommand{\noopsort}[1]{}

\end{document}